\documentclass[12pt]{article}
\usepackage{amsmath}
\usepackage{graphicx,psfrag,epsf}
\usepackage{enumerate}
\usepackage{natbib}
\usepackage{url} % not crucial - just used below for the URL 
\usepackage{multirow}
\usepackage{booktabs}

\usepackage{xcolor}
\usepackage{float}
\usepackage{booktabs,makecell}
\usepackage{amsthm}
\usepackage{amsfonts}
\usepackage{amssymb}
\usepackage{hyperref}
\hypersetup{
     colorlinks   = true,
  linkcolor=blue,
            urlcolor=blue,
            citecolor=blue
}
\usepackage{xr}
\usepackage[ruled,vlined,onelanguage]{algorithm2e}
\newcommand{\blind}{0}
  
\begin{document}

\def\spacingset#1{\renewcommand{\baselinestretch}%
{#1}\small\normalsize} \spacingset{1}

%Own Definition
\def\bmY{\boldsymbol{Y}}
\def\bmW{\boldsymbol{W}}
\def\bmM{\boldsymbol{M}}
\def\bmE{\boldsymbol{\mathcal{E}}}
\def\bmlambda{\boldsymbol{\lambda}}
\def\bmrho{\boldsymbol{\rho}}
\def\Cov{{\rm Cov}}
\def\rmvec{{\rm vec}}
\def\bmSigma{\boldsymbol{\Sigma}}
\def\mbR{\mathbb{R}}
\def\mbW{\mathbb{W}}
\def\bmI{\boldsymbol{I}}
\def\bmy{\boldsymbol{y}}
\def\bmvarepsilon{\boldsymbol{\varepsilon}}
\def\bmxi{\boldsymbol{\xi}}
\def\bmeta{\boldsymbol{\eta}}
\def\lbk{\left\{}
\def\rbk{\right\}}
\def\lak{\left|}
\def\rak{\right|}
\def\lmk{\left[}
\def\rmk{\right]}
\def\lsk{\left(}
\def\rsk{\right)}
\def\bmx{\boldsymbol{x}}
\def\bmz{\boldsymbol{z}}
\def\bmv{\boldsymbol{v}}
\def\bmmu{\boldsymbol{\mu}}
\def\bmzeta{\boldsymbol{\zeta}}
\def\bmsigma{\boldsymbol{\sigma}}
\def\bmbeta{\boldsymbol{\beta}}
\def\var{{\rm var}}
\def\bml{\boldsymbol{1}}
\def\bmo{\boldsymbol{0}}
\def\diag{{\rm diag}}
\def\bmrho{\boldsymbol{\rho}}
\def\E{{\rm E}}
\def\bmgamma{\boldsymbol{\gamma}}
\def\bmalpha{\boldsymbol{\alpha}}
\def\bmvtheta{\boldsymbol{\vartheta}}
\def\laak{\left\|}
\def\raak{\right\|}
\def\OLS{{\rm OLS}}
\def\GLS{{\rm GLS}}
\def\FGLS{{\rm FGLS}}
\def\LS{{\rm LS}}
\def\FLS{{\rm FLS}}
\def\bmtheta{\boldsymbol{\theta}}
\def\tr{{\rm tr}}
\def\mbY{\mathbb{Y}}
\def\mbX{\mathbb{X}}
\def\rmE{{\rm E}}
\def\rmP{{\rm P}}
\def\Cov{{\rm Cov}}
\def\Var{{\rm Var}}
\def\mS{{\mathcal{S}}}
\def\mU{{\mathcal{U}}}
\def\mN{{\mathcal{N}}}
\def\mV{{\mathcal{V}}}
\def\vech{{\rm vech}}

\def\be{\begin{equation}}
\def\ee{\end{equation}} 
\def\ben{\begin{equation*}}
\def\een{\end{equation*}}
\def\bea{\begin{eqnarray}}
\def\eea{\end{eqnarray}}
\def\bda{\begin{eqnarray*}}
\def\eda{\end{eqnarray*}}
\numberwithin{equation}{section}

\newtheorem{theorem}{Theorem}
\newtheorem{proposition}{Proposition}
\newtheorem{corollary}{Corollary}
\newtheorem{assumption}{Assumption}
\newtheorem{lemma}{Lemma}
\newtheorem{example}{Example}
%\newtheorem{algorithm}{Algorithm}

%%%%%%%%%%%%%%%%%%%%%%%%%%%%%%%%%%%%%%%%%%%%%%%%%%%%%%%%%%%%%%%%%%%%%%%%%%%%%%

\if0\blind
{\spacingset{1.5} 
  \title{\bf On the Subbagging Estimation for Massive Data}
  \date{}
  \author{Tao Zou$^a$, Xian Li$^a$, Xuan Liang$^a$ %\thanks{Address for correspondence: Xuan Liang, 26C Kingsley Street, The Australian National University, Acton ACT 2601, Australia.
%E-mail: xuan.liang@anu.edu.au.}\hspace{0.2cm} 
and Hansheng Wang$^b$ \\
    \textit{$^a$The Australian National University and $^b$Peking University}}
  \maketitle

} \fi

\if1\blind
{\spacingset{2} 
  \bigskip
  \bigskip
  \bigskip
  \begin{center}
    {\LARGE\bf On the Subbagging Estimation for Massive Data}
\end{center}
  \medskip
} \fi

\bigskip
\begin{abstract}

\spacingset{1.2}

This article introduces subbagging (subsample aggregating) estimation approaches for big data analysis with memory constraints of computers. Specifically, for the whole dataset with size $N$, $m_N$ subsamples are randomly drawn, and each subsample with a subsample size $k_N\ll N$ to meet the memory constraint is sampled uniformly without replacement. Aggregating the estimators of $m_N$ subsamples can lead to subbagging estimation. To analyze the theoretical properties of the subbagging estimator, we adapt the incomplete $U$-statistics theory with an infinite order kernel to allow overlapping drawn subsamples in the sampling procedure. Utilizing this novel theoretical framework, we demonstrate that via a proper hyperparameter selection of  $k_N$ and $m_N$, the subbagging estimator can achieve $\sqrt{N}$-consistency and asymptotic normality under the condition $(k_Nm_N)/N\to \alpha \in (0,\infty]$. Compared to the full sample estimator, we theoretically show that the $\sqrt{N}$-consistent subbagging estimator has an inflation rate of $1/\alpha$  in its asymptotic variance. Simulation experiments are presented to demonstrate the finite sample performances. An American airline dataset is analyzed to illustrate that the subbagging estimate is numerically close to the full sample estimate, and can be computationally fast under the memory constraint.

\end{abstract}

\noindent%

\vfill

\newpage
\spacingset{1.45} % DON'T change the spacing!

\section{Introduction}
\label{sec:intro}

{\color{black}With the rapid development of information and technology, more and more datasets with massive data sizes are unprecedentedly available for econometric and statistical analysis. Such enormous sizes of data bring in both a blessing and a curse. 
On one hand, the ultra large data size can lead to high estimation efficiency, which means much more accurate estimation output. 
On the other hand, analyzing the immense size of data is possibly beyond the limits of our computers. For example, the massive datasets are sometimes too large to be read into a computer's memory. A personal computer (PC) typically encounters this bottleneck, while a supercomputer with petabytes of data storage also has limited node memory in the process of computation (e.g., the supercomputer Gadi at \url{https://opus.nci.org.au/display/Help/Queue+Limits}).

To alleviate the bottleneck of memory constraints, one possible approach   is to downsize the data volumes by constructing a subset of data randomly, or in other words, drawing a  subsample, instead of analyzing the whole dataset. 
Though the economic wisdom of diminishing returns also suggests that we may not need the entire large dataset  to estimate parameters or perform hypothesis testing, constructing subsamples randomly for massive data analysis have not been sufficiently studied in the econometrics literature except for a brief exposition by \citet{ng2017opportunities}, a seminal review article by \citet{lee2020econometric}, and \citet{lee2020sketching} about two-stage least squares estimation for big data, where these references refer to this approach as the term ``sketching''. Analyzing a sketch/subsample of big data has also drawn the attention of statisticians and machine learning researchers in recent years; see, e.g., \citet{dhillon2013new}, \citet{wang2018optimal} and \citet{yu2020optimal}. A more complete review can be found in \citet{lee2020econometric}.

Except for the memory constraint, \citet{lee2020econometric} also provides other motivations for working with sketches/subsamples of big data, and we summarise those motivations below. First, the massive data are of limited practical use if they are too expensive to store. In this case, a researcher may need to get a sense from a small  subsample of whether the expensive full sample analysis would be worth pursuing. Second, even if a computer's memory is enough to cache the big dataset,  drawing and loading subsamples are much faster than loading the whole big dataset into  the memory. A subsample can be adequate when a researcher is still attempting to specify an appropriate model. Third, the computation for complex estimating problems based on a subsample can be much easier, and debugging in the process is much faster as well. Last, it may only be allowed to circulate a subsample of dataset in real practice due to the reasons of confidentiality.

Though sketching/subsampling can alleviate the computer bottlenecks, the statistical view for efficient estimation and inference  inevitably favors using as many observations as possible.  Specifically, the estimation based on one subsample that can satisfy the memory constraint, is far less efficient than the full sample estimation.  
In order to improve the efficiency, the 
aggregation methods (e.g., bagging), have been used as a general-purpose procedure for reducing the variance in statistical learning \citep{hastie2009elements}.  Similar ideas can be adopted via aggregating more than one subsample, which leads to the subsample aggregating or subbagging  approaches (suggested by, e.g.,  \citealt{andonova2002simple}, \citealt{buhlmann2003bagging} and \citealt{zaman2009effect}, for other machine learning problems). \citet{lee2020econometric} refers to this approach as the term ``combining sketches'' and they find that pooling over subsamples always provides more efficient estimates.  By the principle of divide and conquer, drawing several subsamples and computing their estimates can be realized in parallel, and hence are still computationally inexpensive. Instead of efficiency gains, combining several subsamples which are drawn uniformly at random can help avoid the possible sample bias problem caused by drawing only one subsample.

Different from the existing literature, the focus of this article is on establishing the asymptotic theory for the subbagging estimator/combining sketches, such that the efficiency gain from the subbagging approach can be clearly revealed in the asymptotic variance of the estimator. We specifically  target on a general estimation approach of solving estimating equations.  This general estimation framework comprises  a wide class of estimators, including but not limited to the maximum likelihood, least squares, method of moments and generalized estimating equation estimators as special cases. Since econometric models usually specify a set of moment conditions 
 for the true parameter, the approach of solving estimating equations  is also widely used in the econometrics literature (see, e.g., \citealp{rilstone1996second,kim2016higher}).

 Specifically, for the whole dataset with massive data size $N$, we propose the subbagging estimator which is the aggregation of the subsample estimators computed based on $m_N$ subsamples, respectively. The size of each subsample  $k_N$ is set to be far less than $N$ to meet the memory constraint.  Based on the subbagging literature (see, e.g., 
  \citealt{buhlmann2003bagging} and \citealt{mentch2016quantifying}), the $m_N$ subsamples
are independently and identically drawn from $S_N^{k_N}$ with equal probability $1/{N\choose k_N}$, where $S_N^{k_N}$ is a set consisting of all the $N\choose k_N$ subsamples of data with size $k_N$. This sampling procedure allows us  to directly draw subsamples from the hard drive, which is important in the sense that  it is almost impossible to  load  the terabytes or petabytes of full data into memory for random sampling. Specifically, drawing  each subsample
from $S_N^{k_N}$ with equal probability $1/{N\choose k_N}$ can be accomplished via the  simple random sampling without replacement, which only requires $O(k_N)$ of memory and $O(k_N\log_2 k_N)$ of sampling time  (see, e.g., \citealt{Gupta1984algorithm}). Hence, drawing $m_N$ subsamples independently and identically from the hard drive is  feasible and computationally efficient in real practice.

Based on the sampling procedure, the major challenge that we encounter in establishing the asymptotic theory for the subbagging estimator is that any one subsample (one subset of data) from the drawn $m_N$ subsamples is possibly overlapping with the other subsample (the other subset). The standard independent asymptotics cannot deal with the aggregation of subsamples that possibly have overlap with each other.  To overcome this challenge, we extend the theory of the incomplete $U$-statistics with an infinite order kernel, which was originally developed for the random forests (see, e.g., \citealt{mentch2016quantifying,peng2019asymptotic}), and then we use it to show the theoretical properties of the subbagging estimator.

This novel asymptotic framework leads to three important findings  for the subbagging estimation approach. The first is that   subbagging  can indeed reduce the asymptotic variance of estimators as the number of drawn subsamples $m_N$ increases. In particular, the subbagging estimator can achieve $\sqrt{N}$-consistency and asymptotic normality, but has an inflation rate of $1/\alpha$ in its asymptotic variance compared to the full sample estimator, where $\alpha=\lim (k_Nm_N/N)\in (0,\infty]$. The second is that the subbagging approach cannot improve the order of the asymptotic bias, which relies on the subsample size $k_N$.  As a consequence, in order to achieve $\sqrt{N}$-consistency for the subbagging estimation, the asymptotic bias  needs to be taken into the consideration. To this end, we propose 
  two subbagging algorithms with  proper hyperparameter selection of  $k_N$ and $m_N$,  to balance  the computation cost and   achieve $\sqrt{N}$-consistency at the same time. The third is that  a subbagging variance estimation approach is  proposed in this article for inference of parameters  under the big data setting. This can be treated as extensions of the $m$-out-of-$n$ bootstrap (see, e.g., \citealt{bickel1988richardson,bickel2012resampling}), or the subsampling methods developed for frequentist inference (see, e.g., \citealt{hong2006fast}).

The rest of this paper is organized as follows. Section \ref{sec:ensemble} introduces the subbagging estimator based on  a simple average ensemble and develops its theoretical properties. Section \ref{sec:sbc} proposes bias correction before the simple average ensemble and demonstrates its advantage.  Section \ref{sec:vare} provides a consistent subbagging variance estimation approach for inference of parameters. 
Simulation studies and an application of the American airline dataset are elaborated in Section \ref{sec:n} to illustrate the usefulness of the proposed subbagging methods. Section \ref{sec:co} concludes the paper with a discussion. All theoretical
proofs are relegated to Appendix. Their extensive derivations are presented in the supplementary material.

}

{\color{black}
\section{Subbagging Estimation}

\label{sec:ensemble}

\subsection{Full sample, subsample and subbagging estimators}

Consider that we have observations of $p$-dimensional data vectors $Z_1,\cdots,Z_N\stackrel{iid}\sim Z$ and $Z$ follows a distribution $F_{Z}$. Let $\theta_0$ be the $d$-dimensional true  parameter vector of interest. The aim of this article is to consider the estimation and inference problems for parameter $\theta_0$ under the massive data setting, i.e., the sample size $N$ is extremely large. In the econometrics and statistics literature, various estimation approaches have been developed to estimate the parameter $\theta_0$ based on data $Z_1,\cdots,Z_N$. Without loss of generality, 
we can denote any one of those estimators by $\hat\theta_N(Z_1,\cdots,Z_N)$, which is a function of the full sample $Z_1,\cdots,Z_N$.

The ultra large data size $N$ can lead to high estimation efficiency of the full sample estimator $\hat\theta_N(Z_1,\cdots,Z_N)$, for instance, the usual $\sqrt{N}$-consistency for the estimator.  However, computing $\hat\theta_N(Z_1,\cdots,Z_N)$ requires loading the whole big dataset in  the memory,  which is not plausible under the memory constraint.  In order to satisfy the memory constraint of a certain computer, we possibly can only analyze 
a smaller size $k_N$ of data, where $k_N< N$. Let
$s=\{s_1,\cdots,s_{k_N}\}\subset\{1,\cdots, N\}$ be a subsample (subset) of data $\{1,\cdots,N\}$. In total, we have $N \choose k_N$ subsamples (subsets). We denote $S_{N}^{k_N}$ as a set consisting of all the subsamples with size $k_N$, so that each $s\in  S_N^{k_N}$. Therefore, the size of $S_{N}^{k_N}$, denoted by $|S_{N}^{k_N}|$, is equal to $N \choose k_N$.   
A subsample estimator can be obtained by $\hat\theta_{k_N}(Z_{s_1},\cdots,Z_{s_{k_N}})$, where the estimation function $\hat\theta_{k_N}()$ is the subsample version of $\hat\theta_N()$. If the full sample estimator $\hat\theta_N(Z_1,\cdots,Z_N)$ is $\sqrt{N}$-consistent, then the subsample estimator is $\sqrt{k_N}$-consistent.

Clearly, the rate of the subsample estimator reduces to $\sqrt{k_N}$. Accordingly, even though the subsample estimator requires less memory to compute, the bias and variance of the estimator both increase. In order to reduce the variance, we consider the approach of bootstrap aggregating, or bagging, which is a general-purpose procedure for reducing the variance in  statistical learning \citep{hastie2009elements}.  Here, we adopt similar ideas and consider the subsample aggregating, which can be called subbagging. Ideally we can take all $N\choose {k_N}$ subsamples in $S_{N}^{k_N}$, and build a subbagging estimator
\be\label{eq:comU}
\hat \theta_{k_N,{N\choose {k_N}}}\triangleq\frac{1}{{N\choose {k_N}}}\sum_{s\in S_N^{k_N}}\hat\theta_{k_N}(Z_{s_1},\cdots,Z_{s_{k_N}}), 
\ee
by averaging the $N\choose {k_N}$ subsample estimators.  

Of course, building $N\choose {k_N}$ subsample estimators is computationally infeasible even for  moderate size $N$, and a substantial improvement in computation efficiency can be made by building and taking average over only $m_N<{N\choose {k_N}}$ subsample estimators. In this case, the subbagging estimator becomes
\be\label{eq:emeq}
\hat \theta_{k_N,m_N}=\frac{1}{m_N}\sum_{s\in\mathcal{S}}\hat\theta_{k_N}(Z_{s_1},\cdots,Z_{s_{k_N}}),
\ee
where $|\mathcal{S}|=m_N$. Based on the subbagging literature (see, e.g., 
  \citealt{buhlmann2003bagging} and \citealt{mentch2016quantifying}), we consider that these $m_N$ subsamples in $\mathcal{S}$ 
are independently and identically 
drawn from $S_N^{k_N}$ with equal probability  $1/{N\choose k_N}$. In the following sections, we develop a novel theoretical framework to derive the theoretical properties of the subbagging estimator, and explicitly show how subbagging improves the efficiency of the subsample estimator.

It is worth noting that in the above subbaging approach, one subsample $s^{(1)}=\{s_1^{(1)},\cdots,$ $s_{k_N}^{(1)}\}\in\mathcal{S}$ is possibly overlapping with the other subsample $s^{(2)}=\{s_1^{(2)},\cdots,s_{k_N}^{(2)}\}\in\mathcal{S}$. It is also even possible that $s^{(2)}=s^{(1)}$ due to the sampling strategy of $\mS$.  Therefore, the standard independent asymptotic techniques cannot be directly applied for the statistic (\ref{eq:emeq}).

In fact, the subbagging estimators (\ref{eq:comU}) and (\ref{eq:emeq}), are a complete $U$-statistic and  an incomplete $U$-statistic, respectively, with the kernel function being $\hat\theta_{k_N}()$ and 
the order of kernel  being $k_N$. When $k_N$ is fixed, the resulting incomplete $U$-statistic remains asymptotically normal; 
 see \citet{janson1984asymptotic} or \citet{lee2019u}. However, we require  $k_N\to\infty$ to guarantee $\sqrt{k_N}$-consistency for the convergence of each subsample estimator.  \citet{frees1989infinite} studied the complete $U$-statistic with infinite $k_N$; 
 \citet{mentch2016quantifying} introduced a central limit theorem (CLT) for the incomplete $U$-statistics with  the order of kernel  $k_N\to\infty$; \citet{peng2019asymptotic} further extended the CLT. Nevertheless, \citet{mentch2016quantifying} focused more on solving the inference problem for the random forests, and hence their conditions in the CLT,  including the conditions for $k_N$ and the kernel function,  can be improved for the subbagging estimation problem given here. For instance, they require the subsample size $k_N=o(N^{1/2})$ which will cause a large bias for the subsample estimator.  \citet{peng2019asymptotic} studied the CLT for the random forests under condition $k_N=o(N)$ and some restrictions imposed on the kernel function, but the CLT was established based on a different resampling setting compared to this paper.  
 To this end, we show an improved version of CLT for the general incomplete $U$-statistics with infinite $k_N$ and more general conditions on the kernel function in Lemma \ref{tm:0} of Appendix, where its extensive proof is relegated to the supplementary material. Using Lemma \ref{tm:0}, we obtain the theoretical properties of the subbagging estimator in the following subsection.

\subsection{Asymptotic theory for subbagging estimator}
 \label{sec:Z}
 \label{sec:sub}

 Throughout the following sections, we denote the subsample and full sample estimators by $\hat\theta_{k_N,s}=\hat\theta_{k_N}(Z_{s_1},\cdots,Z_{s_{k_N}})$ and $\hat\theta_N=\hat\theta_N(Z_1,$ $\cdots,Z_N)$, respectively, to simplify the notation.  In this section, we target on two general estimators,
 the $Z$-estimator and $M$-estimator (named by \citealt{van2000asymptotic}), which are widely used in the literature. Without loss of generality, we introduce the $Z$-estimator or the $M$-estimator for a subsample $s=\{s_1,\cdots,s_{k_N}\}$. The full sample estimator can be naturally obtained by letting $k_N=N$ and $s=\{1,\cdots, N\}$.

Specifically, the subsample $Z$-estimator $\hat\theta_{k_N,s}$ satisfies a system of estimating equations
\be\label{eq:estimating}
\sum_{i=1}^{k_N} \psi_{\theta}(Z_{s_{i}})=0
\ee
 with respect to $\theta$, 
where $\psi_\theta(z)$ is a $d$-dimensional measurable function of $z$. 
In contrast, 
the subsample $M$-estimator can be obtained by minimizing an objective function 
$
\sum_{i=1}^{k_N} \mathcal{M}_{\theta}(Z_{s_{i}})
$  with respect to $\theta$, 
where $\mathcal{M}_\theta(z)$ is a $1$-dimensional measurable function of $z$. To unify these two estimation approaches, if we consider $\mathcal{M}_\theta(z)$ to be a convex smooth function with respective to $\theta$, and denote its partial derivative $\partial \mathcal{M}_\theta(z)/\partial \theta$ by $\psi_{\theta}(Z_{s_{i}})$, then the $M$-estimator can also be obtained via solving estimating equations (\ref{eq:estimating}). Accordingly, we consider that the subsample estimator $\hat\theta_{k_N,s}$ solves the estimating equations (\ref{eq:estimating})
in the rest of this paper. It is worth noting that this estimator comprises the maximum likelihood, least squares, method of moments and generalized estimating equation estimators as special cases, and is widely used in the econometrics and statistics literature (see, e.g., \citealp{rilstone1996second,kim2016higher}), since econometric models usually specify a set of moment conditions 
$\rmE \psi_{\theta_0}(Z_i)=0$ for the true parameter $\theta_0$, which is the population version of (\ref{eq:estimating}).

Based on the subsample estimator $\hat\theta_{k_N,s}$, the subbagging estimator $\hat \theta_{k_N,m_N}$ can be obtained via the subsample aggregation in  (\ref{eq:emeq}). In this subsection, we make use of the CLT for the general incomplete $U$-statistics with an infinite order kernel in Lemma \ref{tm:0} of Appendix to derive the asymptotic property of a general subbagging estimator $\hat \theta_{k_N,m_N}$.  For the illustration purpose, we first provide the asymptotic theory for the subbagging mean estimator in the following example.

\begin{example} \label{ex:1}Suppose that we observe the $p$-dimensional data vectors $Z_1,\cdots,Z_N\stackrel{iid}\sim Z$ and $Z$ follows a distribution with mean vector $\mu_0=\rmE(Z)\in\mathbb{R}^p$ and variance-covariance matrix $\Sigma_0=\Var(Z)\in\mathbb{R}^{p\times p}$. The parameter of interest is the multivariate mean vector $\theta_0=\mu_0$, and hence the estimating equations are (\ref{eq:estimating}) with $\psi_{\theta}(Z_{s_{i}})=\psi_{\mu}(Z_{s_{i}})=Z_{s_i}-\mu$. Subsequently, we can obtain the subsample mean estimator $\hat\mu_{k_N,s}=k_N^{-1}\sum_{i=1}^{k_N}Z_{s_i}$ and the subbagging mean estimator $\hat \mu_{k_N,m_N}=m_N^{-1}\sum_{s\in\mS}\hat\mu_{k_N,s}$, both of which are unbiased estimators of $\mu_0$.
\end{example}

 Let $\|\cdot\|_2$ be the Euclidean norm for any generic vector. We then obtain the following theorem for the subbagging mean estimator.

\begin{theorem} \label{tm:0}For Example \ref{ex:1}, assume   $\Sigma_0$ is finite and not singular. Then we have 
\[
\Var\lsk  \mu_{k_N,{m_N}}\rsk=\frac{1}{N}\Sigma_0 \lbk 1+O\lsk \frac{k_N}{N}\rsk\rbk+\frac{1}{k_Nm_N} \Sigma_0\lbk 1-\frac{1}{{N\choose k_N}}\rbk.
\]
In addition,  assume $k_N\to\infty$ and $k_N/N\to 0$ as $N\to\infty$. We then obtain

(1) If  $(k_Nm_N)/N\to \infty$, then $\Var( \mu_{k_N,{m_N}})=\{1+o(1)\}\Sigma_0/N$, and
\[
\frac{  \mu_{k_N,{m_N}}-\mu_0}{\sqrt{1/N}}\stackrel{d}\longrightarrow \mN(0,\Sigma_0).
\]

(2)  If  $(k_Nm_N)/N\to 0$,   then  $\Var( \mu_{k_N,{m_N}})= \{1+o(1)\}\Sigma_0/(k_Nm_N)$. If we further assume $m_N\to\infty$ and $\rmE \|Z\|_2^4<\infty$, then we have 
\[
\frac{  \mu_{k_N,{m_N}}-\mu_0}{\sqrt{1/(k_Nm_N)}}\stackrel{d}\longrightarrow \mN(0,\Sigma_0).
\]

(3) If  $(k_Nm_N)/N\to \alpha \in (0,\infty)$,   then $\Var( \mu_{k_N,{m_N}})=\{1+o(1)\}(1+1/\alpha) \Sigma_0/N$. If we further assume  $\rmE \|Z\|_2^4<\infty$, then we have 
\[
\frac{ \mu_{k_N,{m_N}}-\mu_0}{\sqrt{(1+1/\alpha)/N}}\stackrel{d}\longrightarrow \mN(0,\Sigma_0).
\]

\end{theorem}

Compared to the variance of the subsample estimator $\Var(\hat\mu_{k_N,s})=\Sigma_0/k_N$, Theorem \ref{tm:0} shows that the variance of the subbagging estimator is indeed reduced for all scenarios (1) -- (3). In contrast to the variance of the full sample estimator $\Var(\hat\mu_{N})=\Sigma_0/N$, the variance inflation $ \Sigma_{0}\{ 1-{1}/{{N\choose k_N}}\}/(k_Nm_N)$ in $\Var(  \mu_{k_N,{m_N}})$ is due to  the extra randomness in sampling $s$ from $S_N^{k_N}$. However, this variance inflation can tend to be smaller than $O(N^{-1})$ as $m_N$ gets larger, which means the subbagging estimator can achieve $\sqrt{N}$-consistency for large $m_N$ as shown in scenarios (1) and (3).

Based on  Theorem \ref{tm:0}, in order to result in the asymptotic normality for the subbagging mean estimator, we require the condition $k_N/N\to 0$. If $m_N$ is selected too small such that $m_Nk_N/N\to \alpha\in[0,\infty)$, then we require a further moment condition $\rmE\|Z\|_2^4<\infty$ to obtain the asymptotic normality. Similar theoretical results can be established for a general subbagging estimator $\hat\theta_{k_N,m_N}$ using the CLT for the incomplete $U$-statistics in Lemma \ref{tm:0} of Appendix.  We summarize the asymptotic property of a general  subbagging estimator $\hat\theta_{k_N,m_N}$ based on estimating equations (\ref{eq:estimating}) in the theorem below.

To present the theorem, let $
\Sigma_{\theta_0}=\Var\big( \psi_{\theta_0}(Z)\big)$ be the variance-covariance matrix of $\psi_{\theta_0}(Z)$, $V_{\theta_0}=\rmE\{{\partial \psi_{\theta_0}(Z)}/{\partial \theta^\top}\}$, $V^{(2)}_{\theta_0}=\rmE\{{\partial^2 \psi_{\theta_0}(Z)}/{(\partial \theta^\top\otimes \partial \theta^\top)}\}$, $H_{\theta_0}=\rmE\big[ \psi_{\theta_0}(Z)^\top \otimes  \{{\partial \psi_{\theta_0}(Z) }/{\partial \theta^\top}\} \big]$, $J_{k_N,s}=k_N^{-1/2} \sum_{i=1}^{k_N}\psi_{\theta_0}(Z_{s_i})$, $J^{(1)}_{k_N,s}=k_N^{-1/2}\sum_{i=1}^{k_N}\{\partial \psi_{\theta_0}(Z_{s_i})/$ $\partial \theta^\top-V_{\theta_0}\}$ and 
\be\label{eq:bkn}
\mathcal{B}_{k_N,s}=-V_{\theta_0}^{-1}\lbk- J^{(1)}_{k_N,s} V_{\theta_0}^{-1} J_{k_N,s}+\frac{1}{2} V_{\theta_0}^{(2)} \lsk V_{\theta_0}^{-1} J_{k_N,s}\otimes V_{\theta_0}^{-1} J_{k_N,s}\rsk  \rbk.
\ee
where $\otimes$ denotes the Kronecker product, and if we denote $(i,j)$-th entry of ${\partial \psi_{\theta}(Z)}/{\partial \theta^\top}$ by $\big({\partial \psi_{\theta}(Z)}/{\partial \theta^\top}\big)_{ij}$, then we define
\[
\frac{\partial^2 \psi_{\theta}(Z)}{\partial \theta^\top\otimes \partial \theta^\top}=\lsk \frac{\big({\partial \psi_{\theta}(Z)}/{\partial \theta^\top}\big)_{ij}}{\partial \theta^\top} \rsk.
\]
%Recall  $
%\Sigma_{\theta_0}=\Var\big( \psi_{\theta_0}(Z)\big)$  and $V_{\theta_0}=\rmE\{{\partial \psi_{\theta_0}(Z)}/{\partial \theta^\top}\}$ defined before Theorem \ref{tm:2}, and let  
Also denote $\rmvec(\cdot)$ as the vectorization for any generic matrix.

\begin{theorem}\label{tm:2}  Under Conditions (C1), (C2-I) -- (C2-II) and (C3-I) -- (C3-II)  in Appendix \ref{sec:c}, we obtain the asymptotic expansion 
\[
\hat\theta_{k_N,m_N}-\theta_0=-\frac{1}{\sqrt{k_N}}V_{\theta_0}^{-1}{m_N^{-1}}\sum_{s\in\mS}{J_{k_N,s}}+\frac{1}{k_N}{m_N^{-1}}\sum_{s\in\mS}{\mathcal{B}_{k_N,s}}+O_{\rmP}\lsk \frac{1}{{k_N^{3/2}}}\rsk,
\]
%\[
%{\rm Bias}(\hat\theta_{k_N,m_N})=\frac{1}{k_N} B_{\theta_0}+o\lsk \frac{1}{k_N}\rsk,\textrm{ and }\]
%\[
%\Var(\hat\theta_{k_N,m_N})=V_{\theta_0}^{-1} \Sigma_{\theta_0}\lsk V_{\theta_0}^{-1}\rsk^\top\lmk \frac{1}{N} \lbk 1+O\lsk \frac{k_N}{N}\rsk\rbk+\frac{1}{k_Nm_N}\lbk 1-\frac{1}{{N\choose k_N}}\rbk\rmk \lbk 1+o(1)\rbk,
%\]
as $k_N\to\infty$, where
$
\rmE(m_N^{-1}\sum_{s\in\mS}{\mathcal{B}_{k_N,s}})=-V_{\theta_0}^{-1}\{ -H_{\theta_0}\rmvec( V_{\theta_0}^{-1}) +0.5V^{(2)}_{\theta_0}( V_{\theta_0}^{-1}\otimes V_{\theta_0}^{-1}) \rmvec( \Sigma_{\theta_0})\}$ $\triangleq B_{\theta_0},
$ say, 
 and $m_N^{-1}\sum_{s\in\mS}{\mathcal{B}_{k_N,s}}=O_{\rmP}(1)$. 
In addition,  assume  $k_N/N\to 0$ as $N\to\infty$ and Condition (C4-I) in Appendix \ref{sec:c} holds. We obtain

(i) If  $(k_Nm_N)/N\to \alpha\in (0,\infty]$ and   $k_N/\sqrt{N}\to \infty$, then we have 
\[
\frac{ \hat\theta_{k_N,m_N}-\theta_0}{\sqrt{(1+1/\alpha)/N}}\stackrel{d}\longrightarrow \mN\Big(0,V_{\theta_0}^{-1} \Sigma_{\theta_0}\lsk V_{\theta_0}^{-1}\rsk^\top\Big).
\]

%(1) If  $(k_Nm_N)/N\to \infty$, then $\Var(\hat\theta_{k_N,m_N})=\{1+o(1)\}V_{\theta_0}^{-1} \Sigma_{\theta_0}( V_{\theta_0}^{-1})^\top/N$. If we further assume $k_N/\sqrt{N}\to \infty$, then we have
%\[
%\frac{  \hat\theta_{k_N,m_N}-\theta_0}{\sqrt{1/N}}\stackrel{d}\longrightarrow \mN\Big(0,V_{\theta_0}^{-1} \Sigma_{\theta_0}\lsk V_{\theta_0}^{-1}\rsk^\top\Big).
%\]

(ii)  If  $(k_Nm_N)/N\to 0$,   $m_N\to\infty$ and $k_N/m_N\to \infty$, then we have 
\[
\frac{  \hat\theta_{k_N,m_N}-\theta_0}{\sqrt{1/(k_Nm_N)}}\stackrel{d}\longrightarrow \mN\Big(0,V_{\theta_0}^{-1} \Sigma_{\theta_0}\lsk V_{\theta_0}^{-1}\rsk^\top\Big).
\]

\end{theorem}

It is worth noting that the asymptotic normality is from the leading term $-k_N^{-1/2}V_{\theta_0}^{-1}{m_N^{-1}}$ $\sum_{s\in\mS}{J_{k_N,s}}$ in the asymptotic expansion, which is an incomplete $U$-statistic. Consequently, the CLT for the incomplete $U$-statistics in Lemma \ref{tm:0} of Appendix is used to establish the asymptotic normality in this theorem; see details of the proof in Appendix \ref{sec:proofs}.  Based on  setting (i) in this theorem, we find that  $1/\alpha$ is the variance inflation rate for the subbagging estimator $\hat\theta_{k_N,m_N}$ compared to the full sample {\color{black}estimator} $\hat\theta_{N}$.  If $\alpha=\infty$, define $1/\alpha=0$. In this case,  there is no variance inflation for the subbagging estimator, and its asymptotic variance $N^{-1}V_{\theta_0}^{-1} \Sigma_{\theta_0}( V_{\theta_0}^{-1})^\top$ is exactly equal to that of the full sample estimator $\hat\theta_N$.

Different from Example \ref{ex:1} and Theorem \ref{tm:0}, the general subbagging estimator $\hat\theta_{k_N,m_N}$ in Theorem \ref{tm:2} is a consistent but biased estimator, whose asymptotic bias can be obtained by the mean of $k_N^{-1}{m_N^{-1}}\sum_{s\in\mS}{\mathcal{B}_{k_N,s}}$ in the asymptotic expansion.  Using the results of Theorem \ref{tm:2}, we find that this asymptotic bias is $B_{\theta_0}/k_N=O(1/k_N)$, which is the same as the asymptotic bias of the subsample estimator  $\hat\theta_{k_N,s}$; e.g., see Lemma \ref{pn:1} of Appendix. The reason behind this fact is that solving estimating equations (\ref{eq:estimating}) generally results in a biased estimator. In sum, the subbagging approach can improve the efficiency compared to the subsample estimator as shown in Theorem \ref{tm:0} -- \ref{tm:2}, but cannot improve the bias.

Under setting (i) of  Theorem \ref{tm:2}, the asymptotic variance of the subbagging estimator is $N^{-1}(1+1/\alpha)V_{\theta_0}^{-1} \Sigma_{\theta_0}( V_{\theta_0}^{-1})^\top=O(N^{-1})$, and hence 
the asymptotic mean squared error  is MSE = bias$^2$ + variance = $O(k_N^{-2}+N^{-1})$. In order to make the square of the bias much smaller than the variance, we further require $k_N/\sqrt{N}\to \infty$, which leads to MSE = $O(N^{-1})$, namely the $\sqrt{N}$-consistency for the subbagging estimator. This  gives an intuitive explanation about the reason why the $\sqrt{N}$-consistency requires a technical condition $k_N/\sqrt{N}\to \infty$ in setting (i). Setting (ii) of Theorem \ref{tm:2} seems to be computationally most efficient because $m_N$ only requires a small setting $m_N/(N/k_N)\to 0$. However, the asymptotic variance of $\hat\theta_{k_N,m_N}$ is of order  $(k_Nm_N)^{-1}$ which is far less than $N^{-1}$ compared to the asymptotic variance of the full sample estimator $\hat\theta_{N}$. Consequently, we recommend setting (i) in real practice to balance the computational feasibility and the estimation efficiency.

To illustrate the usefulness of Theorem \ref{tm:2}, we provide the asymptotic theory for the subbagging covariance matrix estimator in the following example.

\begin{example} \label{ex:2}We still consider the setting of Example \ref{ex:1}, however, the parameter of interest is $\theta_0=\big(\mu_0^\top,\vech^\top(\Sigma_0)\big)^\top\in \mathbb{R}^{p+p(p+1)/2}$, where $p(p+1)/2\times 1$ vector $\vech(\Sigma_0)$ only stacks the elements on and below the main diagonal of the $p\times p$ covariance matrix defined in 
\citet[p. 662]{lutkepohl2005new}. Accordingly, the estimating equations are (\ref{eq:estimating}) with $\psi_{\theta}(Z_{s_i})=\begin{pmatrix} Z_{s_i}-\mu\\
\vech\lbk (Z_{s_i}-\mu)(Z_{s_i}-\mu)^\top \rbk-\vech(\Sigma)
\end{pmatrix}$. We subsequently obtain the mean estimators $\hat\mu_{k_N,s}$ and $\hat\mu_{k_N,m_N}$ which are the same as those in Example \ref{ex:1}, 
the subsample covariance matrix estimator $\hat\Sigma_{k_N,s}=k_N^{-1}\sum_{i=1}^{k_N} (Z_{s_i}-\hat\mu_{k_N,s})(Z_{s_i}-\hat\mu_{k_N,s})^\top$ and the subbagging covariance matrix estimator $\hat \Sigma_{k_N,m_N}=m_N^{-1}\sum_{s\in\mS}\hat\Sigma_{k_N,s}$.
\end{example}

Let $K^{(3)}=\rmE\lmk \lbk (Z-\mu_0)\otimes (Z-\mu_0)\rbk (Z-\mu_0)^\top\rmk$, $K^{(4)}=\rmE[\{ (Z-\mu_0)(Z-\mu_0)^\top \} \otimes \{(Z-\mu_0)(Z-\mu_0)^\top \}]$, $L_p$ be the $p(p+1)/2\times p^2$ elimination matrix defined in 
\citet[p. 662]{lutkepohl2005new}, and $K^{(4)}_L=L_p \{K^{(4)}-\rmvec(\Sigma_0)$ $\rmvec^\top(\Sigma_0)\} L_p^\top$. Then we have the following corollary for the subbagging covariance matrix estimator.

\begin{corollary}\label{cy:1}  For $\hat\theta_{k_N,m_N}=\big(\hat\mu_{k_N,m_N}^\top,\vech^\top(\hat\Sigma_{k_N,m_N})\big)^\top$ in Example \ref{ex:2}, assume $k_N\to\infty$, $k_N/N\to 0$ as $N\to\infty$, and $\rmE\|Z\|_2^8<\infty$. In addition, assume   matrix $\begin{pmatrix} \Sigma_0 & \lsk L_p K^{(3)} \rsk^\top\\
L_p K^{(3)} & K^{(4)}_L
\end{pmatrix}$ is not singular. 
We obtain

(i) If  $(k_Nm_N)/N\to \alpha\in (0,\infty]$ and   $k_N/\sqrt{N}\to \infty$, then we have 
\[
\frac{ \hat\theta_{k_N,m_N}-\theta_0}{\sqrt{(1+1/\alpha)/N}}\stackrel{d}\longrightarrow \mN\lsk 0,\begin{pmatrix} \Sigma_0 & \lsk L_p K^{(3)} \rsk^\top\\
L_p K^{(3)} & K^{(4)}_L
\end{pmatrix}\rsk.
\]

%(1) If  $(k_Nm_N)/N\to \infty$, then $\Var(\hat\theta_{k_N,m_N})=\{1+o(1)\}V_{\theta_0}^{-1} \Sigma_{\theta_0}( V_{\theta_0}^{-1})^\top/N$. If we further assume $k_N/\sqrt{N}\to \infty$, then we have
%\[
%\frac{  \hat\theta_{k_N,m_N}-\theta_0}{\sqrt{1/N}}\stackrel{d}\longrightarrow \mN\Big(0,V_{\theta_0}^{-1} \Sigma_{\theta_0}\lsk V_{\theta_0}^{-1}\rsk^\top\Big).
%\]

(ii)  If  $(k_Nm_N)/N\to 0$,   $m_N\to\infty$ and $k_N/m_N\to \infty$, then we have 
\[
\frac{  \hat\theta_{k_N,m_N}-\theta_0}{\sqrt{1/(k_Nm_N)}}\stackrel{d}\longrightarrow \mN\lsk 0,\begin{pmatrix} \Sigma_0 & \lsk L_p K^{(3)} \rsk^\top\\
L_p K^{(3)} & K^{(4)}_L
\end{pmatrix}\rsk.
\]

\end{corollary}

Corollary \ref{cy:1} is a direct result of Theorem \ref{tm:2} based on the setting of Example \ref{ex:2}. It is worth noting that  the subbagging covariance matrix estimator is consistent but biased, because the subsample estimator $\hat\Sigma_{k_N,s}=k_N^{-1}\sum_{i=1}^{k_N} (Z_{s_i}-\hat\mu_{k_N,s})(Z_{s_i}-\hat\mu_{k_N,s})^\top$  is a biased estimator. Similar to the general subbagging estimator in Theorem \ref{tm:2}, this biasedness is the reason why  we require a technical condition $k_N/\sqrt{N}\to \infty$ to achieve $\sqrt{N}$-consistency in setting (i) of Corollary \ref{cy:1}.

\subsection{Hyperparameter selection for $k_N$ and $m_N$}

 It is worth noting that the memory constraint of a computer can be treated as a constraint on the subsample size $k_N$. In order to both satisfy the memory constraint of the computer and achieve $\sqrt{N}$-consistency for the subbagging estimation, the hyperparameters of $k_N$ and $m_N$ need to be chosen properly based on the theoretical results of Theorems \ref{tm:0} -- \ref{tm:1}. We summarize the $\sqrt{N}$-consistent subbagging estimation algorithm with hyperparameter selection
 in the following algorithm.

 \begin{algorithm}[H]

\label{al:1}
\SetAlgoLined
 \begin{minipage}{14cm}
  \caption{Obtaining the $\sqrt{N}$-consistent subbagging estimator $\hat\theta_{k_N,m_N}$ if the memory constraint allows $k_N=O(N^{1/2+\delta_k})$ for some $0<\delta_k<1/2$.}

  1.  Hyperparameter selection:

 \uIf{solving estimating equations (\ref{eq:estimating}) results in a biased estimator,}{
    set $k_N=\lfloor N^{1/2+\delta_k}\rfloor$ and $m_N=\lfloor \alpha N^{1+\delta_m}/k_N\rfloor$ for  any $\delta_m\geq 0$ and any positive constant $\alpha$;    
  }
  \Else{the setting of $k_N$ does not require $k_N/\sqrt{N}\to\infty$, but still $m_N=\lfloor \alpha N^{1+\delta_m}/k_N\rfloor$.
  }

2. Draw $m_N$ subsamples  independently and identically from $S_N^{k_N}$ with equal probability $1/{N\choose k_N}$ and collect them in $\mS$.

3. For each subsample $s\in\mS$, compute the subsample estimator $
\hat \theta_{k_N,s}
$ which solves the estimating equations  (\ref{eq:estimating}).

     \KwResult{$ \hat \theta_{k_N,m_N}=m_N^{-1}\sum_{s\in\mathcal{S}}\hat\theta_{k_N,s}$.}

  \end{minipage}
\end{algorithm}

If the subbagging estimator is biased  (e.g., the estimator in Theorem \ref{tm:2}), $k_N/\sqrt{N}\to\infty$ is required to achieve $\sqrt{N}$-consistency, and the asymptotic bias is of order $O(1/k_N)$. As a consequence, $k_N$ needs to be set as large as possible to reduce the bias. If the subbagging estimator is unbiased (e.g., the subbagging mean estimator in Theorem \ref{tm:0}, and subbagging the ordinary least squares estimation for linear regression), achieving $\sqrt{N}$-consistency does not require $k_N/\sqrt{N}\to\infty$.

 }

From the perspectives of the computation efficiency and estimation efficiency, a small $\delta_m>0$ {\color{black}in the above algorithm} is preferred, because it results in not only a relatively small $m_N$ and but also a small asymptotic variance of $\hat\theta_{k_N,m_N}$, which is exactly equal to that of the full sample {\color{black}estimator} $\hat\theta_N$. Setting $\delta_m=0$ can lead to an even smaller $m_N$ and thus a faster computation speed, but with a sacrifice of the variance inflation rate $1/\alpha$ for the subbagging estimator.

{\color{black}
 
 \subsection{Sampling from hard drive}
 
 \label{sec:strategy}

  Based on the subbagging literature (see, e.g., 
  \citealt{buhlmann2003bagging} and \citealt{mentch2016quantifying}), we draw $m_N$ subsamples independently and identically from $S_{N}^{k_N}$ with equal probability $1/{N\choose k_N}$ 
in Algorithm \ref{al:1}. Each drawn subsample $s=\{s_1,\cdots, s_{k_N}\}$ is a subset of $\{1,\cdots,N\}$ with size $k_N$, and thus $\{s_1,\cdots, s_{k_N}\}$ should be  
 sampled via the simple random sampling (SRS) without replacement. However, due to the memory constraint, subsamples with size $k_N$ need to be drawn directly from the data stored on the hard drive. 
In order to tackle this task, SRS algorithms without replacement from the hard drive need to be investigated. It is worth noting that those algorithms 
have been studied in the literature.   For example, 
\citet{Gupta1984algorithm} proposes an algorithm of SRS without replacement called GSEL, which has the time complexity of $O(k_N\log_2 k_N)$ and requires $O(k_N)$ storage locations. This algorithm meets the memory constraint and hence can realize SRS without replacement from the hard drive for massive data analysis.

}

\section{Subbagging Estimation with Bias Correction}

\label{sec:sbc}

Algorithm \ref{al:1} can {\color{black}result in a $\sqrt{N}$-consistent subbagging estimator} if the memory constraint allows $k_N=O(N^{1/2+\delta_k})$. {\color{black}However i}n real practice, $k_N$ is possibly restricted to be smaller by the memory constraint of a computer, e.g., $k_N=O(N^\delta)$ and $\delta\leq 1/2$. In this case,  $k_N/\sqrt{N}\to \infty$ in Theorem \ref{tm:2} and Algorithm \ref{al:1}  is no longer satisfied{\color{black}, and hence it is possible that Algorithm \ref{al:1} cannot lead to a $\sqrt{N}$-consistent result. It is worth noting that the requirement $k_N/\sqrt{N}\to \infty$ in Theorem \ref{tm:2} and Algorithm \ref{al:1} is 
more to make the square of the bias {\color{black}much} smaller than the variance in the subbagging procedure. Thus}, if we can reduce the bias  {\color{black}when we solve the estimating equations (\ref{eq:estimating}), the requirement for} large $k_N$  can be released.  {\color{black}For example, in the following we reduce the bias of the covariance matrix estimator in Example \ref{ex:2}, and propose a $\sqrt{N}$-consistent subbagging estimator without requiring $k_N/\sqrt{N}\to \infty$.}

{\color{black}
\begin{example} \label{ex:3}We consider the setting of Example \ref{ex:2}, however, we adopt
the unbiased subsample covariance matrix estimator $\hat\Sigma_{k_N,s}^{(bc)}=(k_N-1)^{-1}\sum_{i=1}^{k_N} (Z_{s_i}-\hat\mu_{k_N,s})(Z_{s_i}-\hat\mu_{k_N,s})^\top$, which can be treated as the bias-corrected version of $\hat\Sigma_{k_N,s}$. Subsequently, we propose a new subbagging covariance matrix estimator $\hat \Sigma_{k_N,m_N}^{(bc)}=m_N^{-1}\sum_{s\in\mS}\hat\Sigma_{k_N,s}^{(bc)}$, which is also an unbiased estimator of $\Sigma_0$.
\end{example}

\begin{theorem}\label{cy:2}  Let $\hat\theta_{k_N,m_N}^{(bc)}=\big(\hat\mu_{k_N,m_N}^\top,\vech^\top(\hat\Sigma_{k_N,m_N}^{(bc)})\big)^\top$ based on $\hat\mu_{k_N,m_N}$ in  Example \ref{ex:1} and $\hat\Sigma_{k_N,m_N}^{(bc)}$ in Example \ref{ex:3}. Under the conditions given in Corollary \ref{cy:1}, 
 all the results in Corollary \ref{cy:1} hold for $\hat\theta_{k_N,m_N}^{(bc)}$ without requiring $k_N/\sqrt{N}\to \infty$ in (i) and  $k_N/m_N\to \infty$ in (ii). 
\end{theorem}

Theorem \ref{cy:2} indicates that the bias correction before the subbagging approach can lead to a $\sqrt{N}$-consistency estimator as $k_N$ grows slower than $\sqrt{N}$. This motivates us to consider a similar idea to deal with the general subbagging estimators. Specifically,} 
we {\color{black}first} propose three bias-corrected subsample {\color{black}estimator}s $\hat\theta_{k_N,s}^{(bc1)}$, $\hat\theta_{k_N,s}^{(bc2)}$ and $\hat\theta_{k_N,s}^{(bc3)}${\color{black}, relative to the estimator $\hat\theta_{k_N,s}$ that directly solves the estimating equations (\ref{eq:estimating}).}

The first  bias-corrected subsample {\color{black}estimator} is defined by $\hat\theta_{k_N,s}^{(bc1)}=\hat\theta_{k_N,s}-k_N^{-1}B_{\hat\theta_{k_N,s}},$ 
which is simply the subsample {\color{black}estimator} minus its bias estimation, where $B_{\theta}$ is given {\color{black}in Theorem \ref{tm:2}}. {\color{black}Based on different model settings}, it is possible that $B_{\theta}$ is hard to derive or does not have a closed-form. In this case, we can replace $B_{\theta}$ by its sample version and let 
\[
\widehat V_{k_N,s}(\theta)=k_N^{-1}\sum_{i=1}^{k_N}\frac{\partial \psi_{\theta}(Z_{s_i})}{\partial \theta^\top},\widehat V^{(2)}_{k_N,s}(\theta)=k_N^{-1}\sum_{i=1}^{k_N}\frac{\partial^2 \psi_{\theta_0}(Z_{s_i})}{\partial \theta^\top\otimes \partial \theta^\top},\textrm{ and } 
\]
\ben
\color{black}
\left.\begin{aligned}
\widehat B_{k_N,s}(\theta)=- \widehat V^{-1}_{k_N,s}(\theta)\biggl[&- \frac{1}{k_N}\sum_{i=1}^{k_N}\lbk \frac{\partial \psi_{\theta}(Z_{s_i})}{\partial \theta^\top}-\widehat V_{k_N,s}(\theta)\rbk  \widehat V^{-1}_{k_N,s}(\theta) \psi_{\theta}(Z_{s_i}) \\
&+ \frac{1}{2}\widehat V^{(2)}_{k_N,s}(\theta) \frac{1}{k_N} \sum_{i=1}^{k_N}\lbk \widehat   V^{-1}_{k_N,s}(\theta) \psi_{\theta}(Z_{s_i}) \otimes  \widehat V^{-1}_{k_N,s}(\theta) \psi_{\theta}(Z_{s_i})  \rbk \biggl].
\end{aligned}\right.
\een
\citet{rilstone1996second} and \citet{kim2016higher} have shown $\widehat B_{k_N,s}(\hat\theta_{k_N,s})$ is a consistent estimator of $B_{\theta_0}$. Accordingly, we have the second bias-corrected subsample {\color{black}estimator}  defined by $\hat\theta_{k_N,s}^{(bc2)}=\hat\theta_{k_N,s}-k_N^{-1}\widehat B_{k_N,s}(\hat\theta_{k_N,s}).$ 
Making use of     $\widehat B_{k_N,s}(\theta)$ defined above, \citet{firth1993bias} and \citet{kim2016higher} further proposed a new {\color{black}estimator} $\hat\theta_{k_N,s}^{(bc3)}$ which solves the estimating equations
\[\color{black}\sum_{i=1}^{k_N}\psi_\theta(Z_{s_i})
+\widehat V_{k_N,s}(\theta)\widehat B_{k_N,s}(\theta)=0,
\]
and $\hat\theta_{k_N,s}^{(bc3)}$ is the proposed third bias-corrected subsample {\color{black}estimator}.  

%Let 

%We then have the higher-order expansion for these three estimators in the following proposition.

%It is worth noting that the higher-order expansions are the same for the three bias-corrected subsample {\color{black}estimator}s, and the bias of these subsample estimators reduces to $o(k_N^{-1})$ {\color{black}compared to (\ref{eq:rate2})}. 

Let $\hat\theta_{k_N,s}^{(bc)}$ be any one of the proposed three bias-corrected subsample {\color{black}estimator}s.  Using these estimators, we obtain the bias-corrected subbagging estimator {\color{black}by}
\[\color{black}
\hat \theta_{k_N,m_N}^{(bc)}=\frac{1}{m_N}\sum_{s\in\mathcal{S}}\hat\theta_{k_N,s}^{(bc)}.
\]
%Based on {\color{black}Lemma \ref{pn:2}},  the theoretical properties of this subbagging estimator can be derived from the following expansion
%\be\label{eq:bckm}
%\hat\theta_{k_N,m_N}^{(bc)}-\theta_0=-\frac{1}{k_N}V_{\theta_0}^{-1}\frac{1}{m_N}\sum_{s\in\mathcal{S}}h_{\theta_0}(Z_{s_1},\cdots,Z_{s_{k_N}})+\widetilde \mU_{\theta_0,k_N,m_N}+O_{\rmP}\lsk \frac{1}{{k_N^{3/2}}}\rsk,
%\ee
%where
%\[
%\widetilde \mU_{\theta_0,k_N,m_N}=\frac{1}{m_N}\sum_{s\in\mS}\frac{\mathcal{B}_{k_N,s}-B_{\theta_0}}{k_N}.
%\]
%Note that the first term on the right hand side of equation (\ref{eq:bckm}) is exactly the same as the leading term {\color{black}in (\ref{eq:subbagging})}. 
%Thus it is sufficient to show the rate of $\widetilde \mU_{\theta_0,k_N,m_N}$, which is in fact a new incomplete $U$-statistic with mean zero based on the definition of $\mathcal{B}_{k_N,s}$ in (\ref{eq:bkn}). In the proof of the following theorem, we apply Lemma \ref{tm:0} in Appendix to derive $\Var(\widetilde \mU_{\theta_0,k_N,m_N})$, which can lead to  $\widetilde \mU_{\theta_0,k_N,m_N}=O_{\rmP}(k_N^{-3/2})$. This further implies 
%the following asymptotic results for the bias-corrected subbagging estimator.
{\color{black}The following theorem derives the asymptotic property of a general  bias-corrected subbagging  estimator $\hat\theta_{k_N,m_N}^{(bc)}$.

\begin{theorem}\label{tm:3}  Under Conditions (C1) -- (C4) in Appendix \ref{sec:c}, we obtain the asymptotic expansion 
\[
\hat\theta_{k_N,m_N}-\theta_0=-\frac{1}{\sqrt{k_N}}V_{\theta_0}^{-1}{m_N^{-1}}\sum_{s\in\mS}{J_{k_N,s}}+\frac{1}{k_N}{m_N^{-1}}\sum_{s\in\mS}\lsk {\mathcal{B}_{k_N,s}}-B_{\theta_0}\rsk+O_{\rmP}\lsk \frac{1}{{k_N^{3/2}}}\rsk,
\]
as $k_N\to\infty$, where
$B_{\theta_0}
$ is defined in Theorem \ref{tm:2}. 
In addition,  assume  $k_N/N\to 0$ as $N\to\infty$. We obtain

(i) If  $(k_Nm_N)/N\to \alpha\in (0,\infty]$,    $k_N=O(\sqrt{N})$ and $k_N/\sqrt[3]{N}\to \infty$, then we have 
\[
\frac{ \hat\theta_{k_N,m_N}^{(bc)}-\theta_0}{\sqrt{(1+1/\alpha)/N}}\stackrel{d}\longrightarrow \mN\Big(0,V_{\theta_0}^{-1} \Sigma_{\theta_0}\lsk V_{\theta_0}^{-1}\rsk^\top\Big).
\]  

(ii)  If  $(k_Nm_N)/N\to 0$,    $m_N\to\infty$, $k_N=O(m_N)$, and $k_N/\sqrt{m_N}\to \infty$, then we have 
\[
\frac{  \hat\theta_{k_N,m_N}^{(bc)}-\theta_0}{\sqrt{1/(k_Nm_N)}}\stackrel{d}\longrightarrow \mN\Big(0,V_{\theta_0}^{-1} \Sigma_{\theta_0}\lsk V_{\theta_0}^{-1}\rsk^\top\Big).
\]

\end{theorem}

}

{\color{black} Different from Example \ref{ex:3} and Theorem \ref{cy:2}, the general bias-corrected subbagging estimator $\hat\theta_{k_N,m_N}$ in Theorem \ref{tm:3} is consistent, but  still a biased estimator. However, the asymptotic bias is 
no longer from the mean of $k_N^{-1}{m_N^{-1}}\sum_{s\in\mS}{(\mathcal{B}_{k_N,s}-B_{\theta_0})}$ in the asymptotic expansion, because using the results of Theorem \ref{tm:2}, we obtain $\rmE\{k_N^{-1}{m_N^{-1}}\sum_{s\in\mS}{(\mathcal{B}_{k_N,s}-B_{\theta_0})}\}$ $=0$.
As a consequence of Theorems \ref{cy:2} -- \ref{tm:3}, we recommend the following algorithm  for hyperparameter selection and obtaining a $\sqrt{N}$-consistent subbagging estimator under the more restrict memory constraint $k_N=O(N^{1/3+\delta_k})$ for some $0<\delta_k\leq 1/6$.

 \begin{algorithm}[H]

\label{al:2}
\SetAlgoLined
 \begin{minipage}{14cm}
  \caption{Obtaining $\sqrt{N}$-consistent subbagging estimator $\hat\theta_{k_N,m_N}^{(bc)}$ if the memory constraint allows $k_N=O(N^{1/3+\delta_k})$ for some $0<\delta_k\leq 1/6$.}

  1.  Hyperparameter selection:

 \uIf{the bias correction approach is implemented but still results in a biased estimator,}{
    set $k_N=\lfloor N^{1/3+\delta_k}\rfloor$ and $m_N=\lfloor \alpha N^{1+\delta_m}/k_N\rfloor$ for any $\delta_m\geq 0$ and any positive constant $\alpha$;    
  }
  \Else{the setting of $k_N$ does not require $k_N/\sqrt[3]{N}\to \infty$, but still $m_N=\lfloor \alpha N^{1+\delta_m}/k_N\rfloor$.
  }

2. Draw $m_N$ subsamples  independently and identically from $S_N^{k_N}$ with equal probability $1/{N\choose k_N}$ and collect them in $\mS$.

3. For each subsample $s\in\mS$, compute the bias-corrected subsample estimator $
\hat \theta_{k_N,s}^{(bc)}
$.

     \KwResult{$ \hat \theta_{k_N,m_N}^{(bc)}=m_N^{-1}\sum_{s\in\mathcal{S}}\hat\theta_{k_N,s}^{(bc)}$.}

  \end{minipage}
\end{algorithm}

}

{\color{black}
Compared to Algorithm \ref{al:1}, the subbagging approach with bias correction in Algorithm \ref{al:2} can achieve $\sqrt{N}$-consistency under a smaller $k_N=O(N^{1/3+\delta_k})$. For example, the 12GiB American airline dataset analyzed in Section \ref{sec:real} of this article has $N=$ 118,914,459 observations in total. It suffices to set $k_N=\lfloor N^{1/2+1/1000}\rfloor=$11,109 if we apply Algorithm \ref{al:1}. However, $k_N$ can be reduced to $\lfloor N^{1/3+1/1000}\rfloor=500$ if we apply Algorithm \ref{al:2}.  It is worth noting that the reduction of $k_N$ causes the rise of $m_N$ in Algorithm \ref{al:2}, and hence the computation cost increases.

}

Based on the idea {\color{black}of Algorithm \ref{al:2}}, one can expect that a second-order bias correction or higher-order bias-correction is needed for more strict memory {\color{black}constraints}, but similar bias correction analyses can be made to resolve this issue.

{\color{black}Analogously}, other bias correction methods can be {\color{black}implemented to reduce the bias}, e.g., jackknife and {\color{black}bootstrap}. However, the jackknife or  bootstrap bias correction requires  a further resampling procedure and solving estimating equations of the resampled data, which is computationally more expensive than the aforementioned bias correction methods.

\section{Variance Estimation}

\label{sec:vare}

In order to realize the {\color{black}inference} on parameter $\theta$, we need to estimate $\Xi_{\theta_0}\triangleq V_{\theta_0}^{-1} \Sigma_{\theta_0}( V_{\theta_0}^{-1})^\top$ which is the {\color{black}variance in the asymptotic normal distribution of the full sample estimator and subbagging estimators}. For the full sample {\color{black}estimator} $\hat\theta_N$, {\color{black}$\Xi_{\theta_0}$ can either be estimated} by $ \Xi_{\hat\theta_N}=V_{\hat\theta_N}^{-1} \Sigma_{\hat\theta_N}( V_{\hat\theta_N}^{-1})^\top$ or 
\be\label{eq:xi}\widehat\Xi_N(\hat\theta_N)=
\lbk \frac{1}{N}\sum_{i=1}^{N}\frac{\partial \psi_{\hat\theta_N}(Z_{i})}{\partial \theta^\top}\rbk^{-1}\lbk \frac{1}{N}\sum_{i=1}^{N}\psi_{\hat\theta_N}(Z_{i}) \psi_{\hat\theta_N}(Z_{i})^\top \rbk  \lbk \frac{1}{N}\sum_{i=1}^{N}\frac{\partial \psi_{\hat\theta_N}(Z_{i})^\top}{\partial \theta}\rbk^{-1}.
\ee
The former variance estimator $ \Xi_{\hat\theta_N}$ is possibly not easy to obtain because $\Xi_{\theta}$ can be difficult to derive or does not have a closed-form. 
The latter variance estimator $\widehat\Xi_N(\hat\theta_N)$ requires the use of full data set $\{1,\cdots,N\}$ even if we have already computed $\hat\theta_N$, and hence does not satisfy the memory constraints $k_N=O(N^{1/2+\delta_k})$ and $k_N=O(N^{1/3+\delta_k})$ in Algorithms \ref{al:1} -- \ref{al:2}.

Similar to bootstrap/bagging, subbagging also provides the variance estimation as $m_N\to \infty$. Specifically, we can consider the subbagging variance estimator
\be\label{eq:omega1}
\widehat\Omega_{k_N,m_N}=\frac{1}{m_N}\sum_{s\in\mS}\lsk \hat\theta_{k_N,s}- \hat\theta_{k_N,m_N}\rsk \lsk \hat\theta_{k_N,s}- \hat\theta_{k_N,m_N}\rsk^\top,\textrm{ or }\ee\be \widehat\Omega^{(bc)}_{k_N,m_N}=\frac{1}{m_N}\sum_{s\in\mS}\lsk \hat\theta^{(bc)}_{k_N,s}- \hat\theta^{(bc)}_{k_N,m_N}\rsk \lsk \hat\theta_{k_N,s}^{(bc)}- \hat\theta_{k_N,m_N}^{(bc)}\rsk^\top.
\label{eq:omega2}
\ee
Using the theoretical results in {\color{black}Theorem \ref{tm:2} and Theorem  \ref{tm:3}, along with} the incomplete $U$-statistics theory of Lemma \ref{tm:0} in Appendix, we obtain the consistency of the subbagging variance estimators in the following theorem.

 \begin{theorem}\label{tm:4} Under the conditions given in Theorem \ref{tm:2},  we obtain 
$ k_N \widehat\Omega_{k_N,m_N}\stackrel{\rmP}\longrightarrow \Xi_{\theta_0}$.  
 Under the conditions given in Theorem \ref{tm:3},  we obtain $ k_N \widehat\Omega^{(bc)}_{k_N,m_N}\stackrel{\rmP}\longrightarrow \Xi_{\theta_0}$.

 \end{theorem}

 Compared to the full sample variance estimators $\Xi_{\hat\theta_N}$ and $\widehat\Xi_N(\hat\theta_N)$, there are three advantages of using the subbagging variance estimators $k_N \widehat\Omega_{k_N,m_N}$ and $k_N \widehat\Omega^{(bc)}_{k_N,m_N}$. First, similar to bootstrap, obtaining the subbagging variance estimators 
 does not require the derivation of the variance $\color{black}\Xi_{\theta_0}= V_{\theta_0}^{-1} \Sigma_{\theta_0}( V_{\theta_0}^{-1})^\top$. Second, the subbagging variance estimators can satisfy the memory constraint $k_N=O(N^{1/2+\delta_k})$ or $k_N=O(N^{1/3+\delta_k})$. Third, if we have already computed the subsample {\color{black}estimates as in Algorithms \ref{al:1} -- \ref{al:2}}, 
the  computation of the subbagging variance estimator $k_N \widehat\Omega_{k_N,m_N}$ or $k_N \widehat\Omega^{(bc)}_{k_N,m_N}$ is more efficient than that of the full sample variance estimator $\widehat\Xi_N(\hat\theta_N)$, because computing $k_N \widehat\Omega_{k_N,m_N}$ or $k_N \widehat\Omega^{(bc)}_{k_N,m_N}$ only requires the aggregation of $m_N$ subsample {\color{black}estimate}s but not $N$ data points, and in both Algorithms \ref{al:1} -- \ref{al:2}, $m_N=\lfloor \alpha N^{1+\delta_m}/k_N\rfloor\ll N$. This computation advantage is demonstrated in the real data analysis of Section \ref{sec:real}.

\section{Numerical Studies}

\label{sec:n}

\subsection{Simulation Studies}
\label{sec:simu}

{\color{black}To assess the finite sample performance of our proposed algorithms and theoretical results, we conduct numerical studies}
by simulating data from {\color{black}linear regression and logistic regression. The ordinary least squares estimation of linear regression and the maximum likelihood estimation of logistic regression are both under the estimation framework that we have discussed in Section \ref{sec:Z}. In this section, we only present the results for logistic regression to save space. The results for linear regression and their related discussions are provided in Section \ref{sec:ASimu} of the supplementary material.

Specifically, for each observation $i\in\{1,\cdots,N\}$ from the logistic regression model}, the data are simulated by $Z_i=(Y_i,X_i)$ with a binary response $Y_i$ and $d=2$ exogenous
covariates $X_{i}=(1, X_{i 1})^{\top},$ where $X_{i 1}$ is independently generated from the standard normal distribution. The corresponding repression coefficients of 
$X_{i}$ are set to be $\theta_0=(\theta_{10},\theta_{20})^\top=(0,1)^{\top} .$ The binary response $Y_{i}$ is generated 
from a Bernoulli distribution with the probability given by 
\[
P\left(Y_{i}=1 | X_{i}\right)=\frac{\exp \left(X_{i}^{\top} \theta_0\right)}{1+\exp \left(X_{i}^{\top} \theta_0\right)}.
\]
We consider three sample sizes $N=$ 2,000, 10,000, and 50,000. {\color{black}Since the maximum likelihood estimator of logistic regression is a biased estimator,
to} satisfy the requirements of Algorithm \ref{al:1} and Algorithm \ref{al:2}, we consider four subsample sizes $k_N=\lfloor N^{5/12}\rfloor$, $\lfloor N^{6/12}\rfloor$, $\lfloor N^{7/12}\rfloor$ and  $\lfloor N^{8/12}\rfloor$, and two settings of $m_N=\lfloor \alpha N/k_N\rfloor$ and $\lfloor \alpha N^{4/3}/k_N\rfloor$  with  $\alpha=1$ and $1/3$.

For each setting, all simulation results are obtained via 1,000 realizations. 
To evaluate the performance of parameter estimates, we define $\hat\theta^{(r)}=(\hat\theta_1^{(r)},\hat\theta_2^{(r)})^\top$ as the vector estimate of $\theta$ obtained via the full sample {\color{black}estimation} or  Algorithm \ref{al:1} or Algorithm \ref{al:2} in the $r$-th realization. For each component of $\theta$, say $\theta_j$, the averaged bias of $\hat\theta_j^{(r)}$, $r=1,\cdots,1000$, is BIAS $=1000^{-1}\sum_{r}( \hat{\theta}_{j}^{(r)}-{\theta}_{j0})$,  and the standard deviation of $\hat\theta_j^{(r)}$ is SD $=\{1000^{-1}\sum_{r}(\hat{\theta}_{j}^{(r)}-1000^{-1}{\sum_{r}\hat{\theta}_{j}^{(r)}})^2\}^{1/2}$.  So the root mean squared error is RMSE $=\sqrt{\textrm{BIAS}^2+\textrm{SD}^2}$. For any generic square matrix $G$, let $(G)_{jj}$ be the $j$-th diagonal element of $G$.
The   asymptotic standard deviation (ASD) of $\hat \theta_j^{(r)}$ is  approximated by $1000^{-1}\sum_r\sqrt{\big(N^{-1}\widehat \Xi_N^{(r)} (\theta_0)\big)_{jj}}$ and the averaged subbagging standard error  (SSE) is $ 1000^{-1}\sum_r\sqrt{\big(N^{-1}k_N\widehat\Omega_{k_N,m_N}^{(r)}\big)_{jj}}$, where $\widehat \Xi_N^{(r)} (\cdot)$ and $\widehat\Omega_{k_N,m_N}^{(r)}$ are those defined in (\ref{eq:xi}), (\ref{eq:omega1}) and (\ref{eq:omega2}) but for the $r$-th realization. Due to the variance inflation of the subbagging estimators, we further define the $\alpha$-adjusted ASD and $\alpha$-adjusted SSE by $\sqrt{1+1/\alpha}$ times  ASD and  SSE, respectively. The SSE and $\alpha$-adjusted SSE, together with the asymptotic normality of the subbagging estimators, lead to  95\% confidence intervals    and $\alpha$-adjusted confidence intervals of  ${\theta}_{j}$, respectively. If we denote either of the confidence intervals as CI$_j^{(r)}$, the empirical coverage probability for CI$_j^{(r)}$ is $\mathrm{CP}=1000^{-1} \sum_{r} I_{\{\theta_{j0}  \in \mathrm{CI}_j^{(r)}\}}$, where $I_{\{\cdot\}}$ is an indicator function.

Table \ref{tb:1}  presents the BIAS, SD and RMSE of the full sample {\color{black}estimate}s with three sample sizes. We also report their memory usage (MEMORY)
for programming in Python by utilizing an Intel Core i7 CPU (3.2GHz) with 32GB of 2667MHz DDR4 memory of a PC as a benchmark. To compare with the proposed subbagging estimates, Tables  \ref{tb:2} -- \ref{tb:4} report the aforementioned performance measures for   $\hat \theta_{k_N, m_N}$ in Algorithm \ref{al:1} and  $\hat \theta_{k_N, m_N}^{(bc2)}$, $\hat \theta_{k_N, m_N}^{(bc3)}$ in Algorithm \ref{al:2}, respectively,  by setting $\alpha=1$. As the logistic regression does not have the closed-form for the first bias-corrected subsample {\color{black}estimator} $\hat\theta_{k_N,s}^{(bc1)}$ introduced in Section \ref{sec:sbc},  we only report the performance measures of subbagging estimates established based on $\hat\theta_{k_N,s}^{(bc2)}$ and $\hat\theta_{k_N,s}^{(bc3)}$. The results for  $\alpha=1/3$ yield similar findings and are presented in Tables \ref{tb:s1} -- \ref{tb:s3} of  the supplementary material to save space.

We obtain four interesting findings if the settings of $k_N$ satisfy the requirements of Algorithms \ref{al:1} -- \ref{al:2} (settings of $k_N$ in bold in Tables \ref{tb:2} -- \ref{tb:4}{\color{black}, which are due to the biasedness of the maximum likelihood estimation for logistic regression}), respectively. The first is that the SD and the RMSE of the subbagging estimates are close to those of the full sample estimates when $N$ is large. This finding indicates the $\sqrt{N}$-consistency of the subbagging estimators. The second is that the subbagging estimates have larger BIAS compared to the full sample estimates, and 
larger $k_N$ can reduce the BIAS.  The third is that the larger setting of $m_N=\lfloor \alpha N^{4/3}/k_N\rfloor$ leads to a better efficiency but not a smaller BIAS, and under this setting, the SD is almost identical to the ASD and SSE. However, the SD for the smaller setting of $m_N=\lfloor \alpha N/k_N\rfloor$  is much closer to the $\alpha$-adjusted ASD and $\alpha$-adjusted SSE, because the variance inflation rate $1/\alpha$ should be taken into the consideration as $k_Nm_N/N\to \alpha\in(0,\infty)$. The corresponding CP performances in Tables \ref{tb:2} -- \ref{tb:4} further confirm this  result. 
These second and third findings  are consistent with {\color{black}Theorem \ref{tm:2} and Theorem  \ref{tm:3}}, which have shown that the bias of the subbagging estimators is determined by $k_N$ only but not $N$ and $m_N$, and the efficiency of the subbagging estimators can be improved by increasing $m_N$.  Lastly, the memory usage of the 
subbagging  estimates is significantly smaller than that of the full sample estimates, and smaller $k_N$ indeed leads to a smaller memory usage.

We also investigate the performance of the subbagging estimates if the settings of $k_N$ do not satisfy the requirements of Algorithms \ref{al:1} -- \ref{al:2}. Table \ref{tb:2} obtained by Algorithm \ref{al:1} shows that the ASD, SSE, CP and their adjusted versions all break down if a smaller $k_N=O(N^{1/2})$ is set. In contrast, 
Tables \ref{tb:3} -- \ref{tb:4} obtained by Algorithm \ref{al:2} have a significant reduce in BIAS under the same $k_N=O(N^{1/2})$ setting as $N$ is large. The corresponding estimates, their standard errors, and empirical coverages also perform well.  Consequently, we recommend using Algorithm \ref{al:2} in practice when the memory of a computer is  not large enough for Algorithm \ref{al:1}.

%To evaluate the computation efficiency of the sampling procedure in Algorithm \ref{al:3}, we follow the same setting as in the above example, and compare our proposed algorithm to the simple random sampling (SRS) with replacement of size $k_N$ in 
%Table \ref{tb:s4} of the supplementary  material. Two important findings are below. (i) As $N$ increases or $k_N$ decreases, the percentage of distinct elements in $(s_1',\cdots,s_{k_N}')$ from Algorithm \ref{al:3} rises  to almost 100\%, and $T_N$ is gradually closer to $k_N$ with standard deviation tending to zero. This finding is consistent with the theoretical results in Proposition \ref{tm:sample}.  (ii) Algorithm \ref{al:3} takes similar time to obtain a subsample of size $k_N$ compared to SRS with replacement of the same size. Both of these findings   show that the computation  of Algorithm \ref{al:3} is as efficient as SRS with replacement under $k_N/N\to 0$ for massive data. It is worth noting that the SRS with replacement and Algorithm \ref{al:3} in this simulation study are both realized via the assistance of the Python package ``CluBear'' (\url{https://pypi.org/project/clubear/0.0.12/}),  which executes highly efficient SRS with replacement directly from the hard drive of a computer. We also use this package to analyze a real dataset below.

\subsection{Real Data Analysis}

\label{sec:real}

For {\color{black}the illustration purpose}, we demonstrate the application of our proposed subbagging methods on   {\color{black}a 12GiB} American airline dataset (\url{https://doi.org/10.7910/DVN/HG7NV7}){\color{black}, although our approach can definitely be applied to analyze terabytes of data in real practice. The dataset} contains the flight arrival and departure information for all commercial flights from 1987 to 2008 in US. {\color{black}The full sample size of data is  $N=$ 118,914,459.} Each sample in the dataset corresponds to one flight record. %,
%consisting of the delayed status (Delayed), the actual departure time 
%(DepTime), the scheduled  departure time,  the actual arrival time,  the scheduled  arrival time (CRSArrTime), the actual elapsed time (ActualElapsedTime), the flying distance (Distance), etc, for this flight. 

The research target of this dataset is to investigate factors that can influence the delayed status of a flight. To this end, we obtain the variable ``Delayed", indicating whether or not the flight is delayed for arrival, as the response variable, and consider three  variables as covariates, including the year of the flight (Year), the scheduled departure time 
(CRSDepTime, in HHmm), and the actual elapsed time (ActualElapsedTime, in minutes), to predict the response variable via logistic regression. It is worth noting that the {\color{black}12GiB} data contain  more variables of flight records. However, for the sake of exogeneity and interpretation  purposes, we select these three covariates. We further scale the covariates  by $1/1000$ before the model fitting. The intercept and the three variables lead to the parameter estimates of their regression coefficients $\theta=(\theta_1,\cdots,\theta_4)^\top$.

\begin{figure}[H]
\centering
  \includegraphics[scale=0.5]{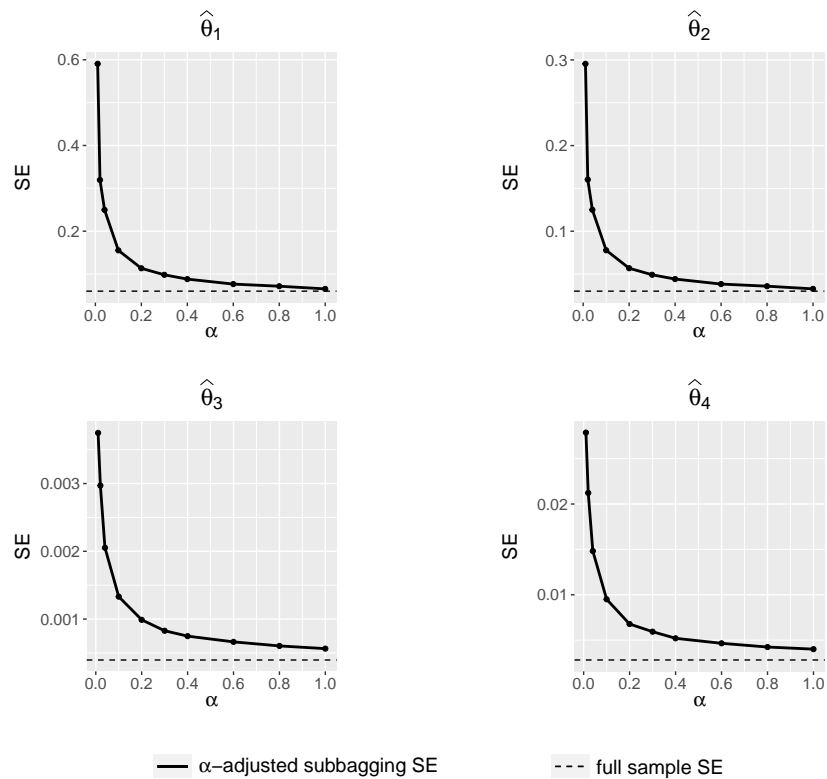}

  \caption{Subbagging standard errors (SSE) under different $\alpha$s for Algorithm \ref{al:1}.    }  \label{fig:1}
\end{figure}

To apply our subbagging algorithms to estimate the {\color{black}logistic} regression coefficients, we consider $k_N=\lfloor N^{1/2+1/1000}\rfloor=$11,109 and $k_N=\lfloor N^{1/3+1/1000}\rfloor=500$ along with $m_N=\lfloor \alpha N/k_N\rfloor$ to balance the computation efficiency and  satisfy the minimum requirements  of Algorithms \ref{al:1} -- \ref{al:2},  respectively. Similar to simulation in Section \ref{sec:simu}, we only  report the results of  the two bias-corrected  subbagging {\color{black}estimate}s $\hat\theta_{k_N,m_N}^{(bc2)}$ and $\hat\theta_{k_N,m_N}^{(bc3)}$ for Algorithm \ref{al:2}.  Further, various settings of $\alpha=0.01$, 0.02, 0.04, 0.1, 0.2, 0.3, 0.4, 0.6, 0.8 and 1 are investigated. To save space, we only report the parameter estimates under settings $\alpha=0.01$ and 0.2 in Table \ref{tb:5}, respectively. However, the $\alpha$-adjusted subbagging standard errors ($\alpha$-adjusted SSE, namely $\sqrt{1+1/\alpha}\times\sqrt{\big(N^{-1}k_N\widehat\Omega_{k_N,m_N}\big)_{jj}}$ used in Section \ref{sec:simu}) for all  $\alpha$s are reported in Figure \ref{fig:1}.

For the illustration purpose, we program in Python via using an Intel Core i7 CPU (3.2GHz) with 32GB of 2667MHz DDR4 memory of a PC, which allows us to load the {\color{black}12GiB} data in the memory, obtain the response and three covariates, and  compute the full sample {\color{black}estimate only for the purpose of} comparison.    The loading of the {\color{black}full} dataset takes {\color{black}27.7 minutes} shown in Table  \ref{tb:5}, while our proposed subbagging algorithms only require less than or equal to {\color{black}6.5} minutes for sampling from data on the hard {\color{black}drive as} $\alpha=0.01$. {\color{black} This fact demonstrates the advantage of the proposed subbagging approaches under smaller $\alpha$ even if the memory of a computer is enough to load the whole dataset in. It is worth noting that the reported loading time for the whole dataset refers to the time that it takes to read the CSV file into a ``NumPy'' array in Python. To accomplish this task, we apply the commonly used function ``pandas.read\_csv(CSV file).values'' from the ``Pandas'' package. Note that compared to  ``read.csv'' function in R, the loading time of using ``Pandas'' package in Python has already been improved (around 31 minutes if loading in R), not to mention that using other packages or functions in Python (e.g., ``NumPy'' function ``numpy.genfromtxt'') can cause a loading failure for the 12 GiB full dataset.}

We also compare the execution times for computing estimates (Estimation Time) and computing SEs (SE Time) in Table  \ref{tb:5}. The SE is  the standard error for the full sample estimate or the $\alpha$-adjusted SSE for the subbagging estimate, and the SE Time is measured after the estimation is completed and saved in the memory. The results in Table  \ref{tb:5} show that the full sample SE takes 1.7 seconds to compute, while Algorithms \ref{al:1} -- \ref{al:2} compute SEs within 0.002 seconds for all settings. Further, the Estimation Times of  $\hat \theta_{k_N,m_N}$ in Algorithm \ref{al:1} and $\hat \theta_{k_N,m_N}^{(bc2)}$, $\hat \theta_{k_N,m_N}^{(bc3)}$ in Algorithm \ref{al:2} under $\alpha=0.01$ are {\color{black}0.4, 2.0 and 3.3} minutes, respectively, much smaller than the {\color{black}20.9} minutes used for computing the full sample estimate. Note that it takes longer time in accomplishing Algorithm \ref{al:2} under $\alpha=0.2$. This is because that the sacrifice of the memory usage (MEMORY in Table \ref{tb:5})  for Algorithm \ref{al:2}  
causes the increase of $m_N=\lfloor \alpha N/k_N\rfloor$  (from $m_N=$ 2,140 in Algorithm \ref{al:1}  to $m_N=47,565$ in Algorithm \ref{al:2} under $\alpha=0.2$), which corresponds to 
larger computational complexity.  {\color{black}But if parallel computing is considered}, since both Algorithms \ref{al:1} -- \ref{al:2} can {\color{black} actually be implemented in parallel},  the computation time for larger $\alpha$ can be  reduced to Table \ref{tb:5}'s  time   divided by the number of parallel {\color{black}processes}. 
It is also noted that all the memory usage of Algorithms \ref{al:1} -- \ref{al:2} is much smaller, i.e., only {\color{black}2666.1KiB $=$ 2.6MiB} and {\color{black}120.0KiB}, respectively, compared to the memory usage of the full sample estimation, {\color{black}12686071.1KiB $=$ 12.1GiB}.  Therefore, our proposed algorithms can be used to meet various memory constraints of computers in real practice.

Table \ref{tb:5} further shows two interesting findings in parameter estimates and SEs. Firstly, the estimates obtained by Algorithm \ref{al:1} -- \ref{al:2} are both numerically close to the full sample estimates.  %compared to those obtained by Algorithm . This is because the sacrifice of the memory usage of Algorithm \ref{al:2} restricts the subsample size of $k_N$.  Based on Theorem \ref{tm:3}, a larger bias of  $O(k_N^{-3/2})$ for the subbagging estimate is generated even after bias correction if $k_N$ is small. 
Secondly, the SEs for the full sample estimate and the subbagging estimates have the same order of magnitude under $\alpha=0.2$, and the SEs for the subbagging estimates under $\alpha=0.01$ are slightly larger. This finding is not surprising since all of the estimators considered in Table \ref{tb:5} are $\sqrt{N}$-consistent, but due to the variance inflation rate $1/\alpha$ for the subbagging estimator, smaller $\alpha$ can result in  a larger variance.  Figure \ref{fig:1} further compares the SEs for the subbagging estimates under larger $\alpha$s  and the SE for the full sample estimate. It is found in Figure \ref{fig:1} that they are almost identical as $\alpha=1$.

To investigate the trade-off between the estimation efficiency  and  the computation efficiency  of the proposed subbagging estimator, we report the $\alpha$-adjusted SSE versus $\alpha$ in Figure \ref{fig:1} for parameter estimates $\hat\theta=(\hat\theta_1,\cdots,\hat\theta_4)^\top$ obtained by  Algorithm \ref{al:1}.  Note that larger $\alpha$ is equivalent to larger $m_N=\lfloor \alpha N/k_N\rfloor$, which is less computationally efficient, but more efficient in estimation. Based on Figure \ref{fig:1}, if one would like to obtain a subbagging estimate whose SE is as close as that of the full sample estimate, then a larger $\alpha$ is preferred. However, in order to balance the computation cost, we can find in Figure \ref{fig:1} that when $\alpha>0.2$, the reduction of SE is quite limited. {\color{black}As a consequence, we recommend using $\alpha=0.2$}. Similar analyses can be made in practice for selecting an optimal $\alpha$ via a ``scree plot'' of Figure \ref{fig:1}.

\begin{figure}[H]
\centering
  \includegraphics[scale=0.5]{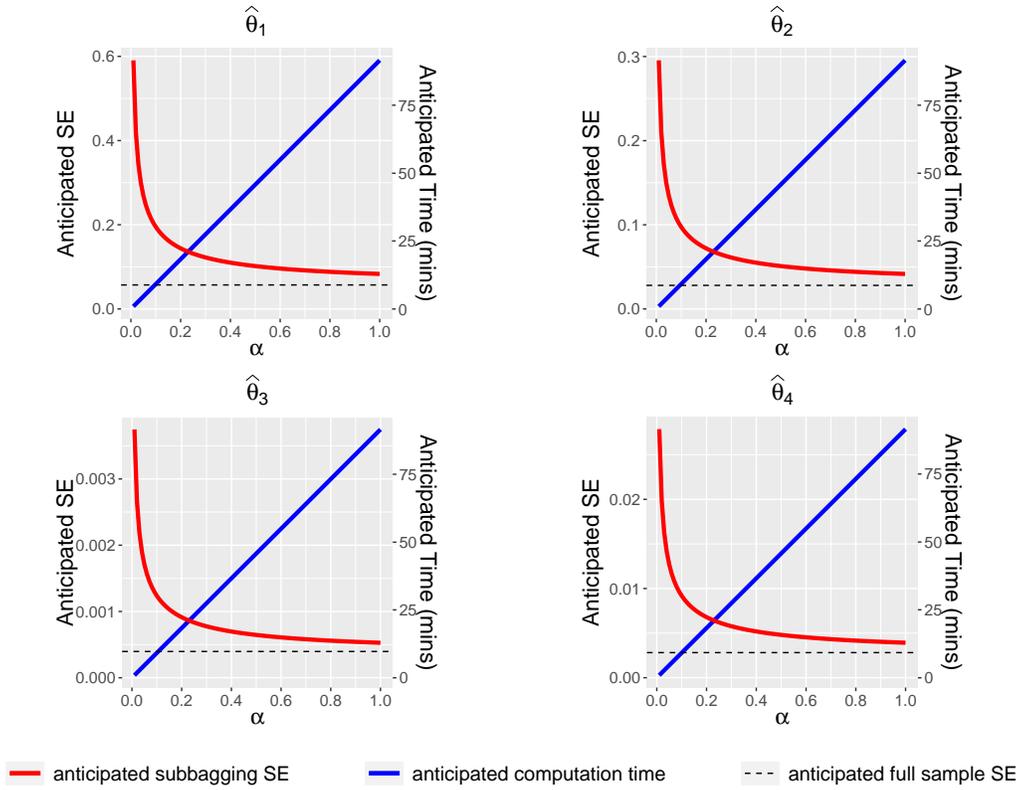}

  \caption{Anticipated SEs and time for Algorithm \ref{al:1}.    }  \label{fig:2}
\end{figure}

 It is worth noting that the $\alpha$-adjusted SSE for the estimate of $\theta_j$ ($j=1,\cdots,4$) in Figure \ref{fig:1}, namely $\sqrt{1+1/\alpha}\times\sqrt{\big(N^{-1}k_N\widehat\Omega_{k_N,m_N}\big)_{jj}}$,  is computed based on a given  $m_N=\lfloor \alpha N/k_N\rfloor$, which requires a longer time to compute for larger $\alpha$ (e.g., see Table \ref{tb:5} given $\alpha=0.2$). So  Figure \ref{fig:1} is not convenient enough for users to fast select an optimal $\alpha$ in real  practice. Note that using {\color{black}Theorem \ref{tm:4}}, we have that $\sqrt{1+1/\alpha}\times\sqrt{\big(N^{-1}k_N\widehat\Omega_{k_N,\lfloor 0.01 N/k_N\rfloor}\big)_{jj}}$ is also a consistent SE for the subbagging estimate, and $\sqrt{\big(N^{-1}k_N\widehat\Omega_{k_N,\lfloor 0.01 N/k_N\rfloor}\big)_{jj}}$ is a consistent SE for the full sample {\color{black}estimate}. We call these standard errors anticipated SEs in 
Figure \ref{fig:2}. The advantage of reporting the anticipated SEs is that we only need to draw a small number of subsamples 
($m_N=\lfloor 0.01 N/k_N\rfloor$) to compute those SEs, and hence they can be obtained within a short period of time ({\color{black}0.9} minutes only to {\color{black}generate} Figure \ref{fig:2}).

{\color{black}In addition},  we are capable of anticipating the computation time of using a PC for any subbagging estimation algorithms  given $\alpha$ ($m_N=\lfloor \alpha N/k_N\rfloor$),  as long as we know the actual computation time given $\alpha=0.01$ ($m_N=\lfloor 0.01 N/k_N\rfloor$). The idea behind it is that 
 the number of drawn subsamples $m_N$, which determines the computational complexity of the algorithms,  increases linearly with respect to $\alpha$. Accordingly,  in Figure \ref{fig:2} we report {\color{black}the} anticipated computation time (the sum of Loading Time, Estimation Time and SE Time)  for Algorithm \ref{al:1} given $\alpha$,  which 
  is  equal to $\alpha/0.01$ times the actual computation time given $\alpha=0.01$.  {\color{black}Accordingly}, Figure \ref{fig:2} reveals how much time it will take to reach some certain level of estimation efficiency for the proposed subbagging estimation.

In summary, 
Table \ref{tb:5} given $\alpha=0.01$  provides a fast {\color{black}sketch on} the big dataset with ensuring  $\sqrt{N}$-consistency for the estimation, while Figure \ref{fig:2} can help users make decisions on whether or not it is worth taking more time or drawing more subsamples in parallel to achieve higher estimation efficiency. %Note that drawing  more subsamples can be easily realized in parallel, and hence the computation time for larger $\alpha$ can be significantly reduced via parallel computing systems. 
All of the above findings demonstrate the usefulness of the subbagging methods for {\color{black}econometric and statistical analysis} on massive data.

\section{Conclusion}

\label{sec:co}

In this article, in order to meet different memory constraints under the massive data setting, 
we introduce two computationally efficient subbagging {\color{black}Algorithms \ref{al:1} -- \ref{al:2} }. Both subbagging algorithms lead to $\sqrt{N}$-consistent estimators that have the same order of estimation efficiency as the full sample estimator. {\color{black}By the principle of divide and conquer, both Algorithms \ref{al:1} -- \ref{al:2} can be realized in parallel, and hence are still computationally inexpensive.} 
 The first algorithm uses the simple average subbagging but {\color{black}generally} requires a larger memory of $O(N^{1/2+\delta_k})$ for computation, where $\delta_k>0$. The second algorithm adopts the bias-corrected subbagging method and can satisfy a more strict memory constraint $O(N^{1/3+\delta_k})$. Obtaining the limiting distributions of the resulting estimators based on these two algorithms is quite challenging due to the fact that one subsample can have overlap with the other subsample in ensembles; see detailed derivations via utilizing the incomplete $U$-statistics theory in the supplementary material. 
We subsequently obtain that both algorithms provide asymptotically normal estimators but with a variance inflation compared to the full sample estimator. These properties allow us to {\color{black}perform} hypothesis testing and {\color{black}construct confidence intervals.} %To further reduce the computational cost, we propose the third algorithm to draw subsamples from the hard drive and provide its theoretical support. 
The performances of the proposed algorithms and the corresponding {\color{black}theoretical} properties are supported by both simulation studies and a real American airline dataset analysis.

We conclude this work by identifying two potential avenues for future research. First, the first-order bias-correction method proposed in this paper can be extended to second-order or higher-order bias-correction, which is expected to satisfy more strict memory constraints.  Second, {\color{black}though the proposed subbagging approach has been analyzed under a general estimation framework of solving estimating equations in this article, other complex estimating problems can also be discussed, especially when the data have a probabilistic structure. In this case, having more data (considering a larger subsample size or drawing more subsamples) is not necessarily more desirable \citep{boivin2006more}}. We believe these extensions would further strengthen the usefulness of the subbagging methods for big data analysis.

\section*{Appendix}

\renewcommand{\theequation}{A.\arabic{equation}}
\setcounter{equation}{0}

\renewcommand{\thesubsection}{A.\arabic{subsection}}

This appendix includes three parts: Appendix \ref{sec:c} introduces technical conditions; Appendix \ref{sec:lemma} provides important lemmas directly used in the proofs of theorems; and Appendix \ref{sec:proofs} presents the proofs of  {\color{black}Theorems \ref{tm:0}} -- \ref{tm:4}, where the proofs of {\color{black}Lemmas \ref{tm:0} -- \ref{la:bs} and Corollary \ref{cy:1}} are relegated to the supplementary material. Throughout this appendix, 
let $\nabla^{\kappa}\psi_\theta(z)=\partial ^\kappa \psi_\theta(z)/(\partial \theta^\top\otimes \cdots\otimes \partial \theta^\top)$ for $\kappa$ being non-negative integers and define $\nabla^{0}\psi_\theta(z)= \psi_\theta(z)$. Let $\|\cdot\|_2$ denote the vector $2$-norm or the matrix $2$-norm. In other words, for any generic vector $x=(x_1,\cdots,x_q)^\top\in\mathbb{R}^q$, $\|x\|_2=(\sum_{i=1}^q |x_i|^2)^{1/2}$, and, for any generic matrix $G\in\mathbb{R}^{m\times q}$,
$
\laak G\raak_2=\sup\{ {\laak Gx\raak_2}/{\laak x\raak_2}:x\in\mathbb{R}^{q}\textrm{ and }x\neq 0\}.
$
%Moreover,  define the element-wise $\ell_{\infty}$ norm for any generic matrix $G$ as $|G|_{\infty}=\|\textrm{vec}(G)\|_{\infty}$, where $\textrm{vec}(G)$ denotes the vectorization for any generic matrix $G$. 
%In addition, we
%denote the Frobenius norm of any generic matrix $G$ as $\|G\|_{F}=\|\textrm{vec}(G)\|_{2}$.

\subsection{Technical Conditions}
\label{sec:c}

We introduce the following technical conditions. 

\noindent (C1) Assume that the true parameter value $\theta_0$ is an interior point of the compact parameter space $\Theta\subset\mathbb{R}^d$ and $\theta_0$ is the unique root of the system of equations $\rmE \psi_\theta(Z)=0$.

\noindent (C2) {\color{black} Assume the following conditions hold:

\begin{enumerate}[align=parleft]

\item[(C2-I)] $z\mapsto \psi_\theta(z)$ is measurable  given any $\theta\in\Theta$;

\item[(C2-II)]  $\theta\mapsto \psi_\theta(z)$ is three times continuously differentiable in $\Theta$  for $P_Z$-almost every $z$, where $P_Z=\rmP\circ Z^{-1}$ is the induced measure by the random vector $Z$;

\item[(C2-III)] $\theta\mapsto \psi_\theta(z)$ is four times continuously differentiable in $\Theta$  for $P_Z$-almost every $z$.

\end{enumerate}

}

{\color{black}
\noindent (C3) Assume the following conditions hold:

\begin{enumerate}[align=parleft]

\item[(C3-I)] $\rmE \sup_{\theta\in \Theta} \|\nabla^{\kappa}\psi_\theta(Z)\|_2^2<\infty,$  for $\kappa=0,1,2,3$;

\item[(C3-II)]  $
\Sigma_{\theta_0}=\Var\big( \psi_{\theta_0}(Z)\big)$  and $V_{\theta_0}=\rmE\{\nabla \psi_{\theta_0}(Z)\}$  are not singular;

\item[(C3-III)] $\rmE \sup_{\theta\in \Theta} \|\nabla^{4}\psi_\theta(Z)\|_2^2<\infty$.

\end{enumerate}

\noindent (C4) Assume the following conditions hold:

\begin{enumerate}[align=parleft]

\item[(C4-I)] $\rmE  \|\psi_{\theta_0}(Z)\|^4_2<\infty$;

\item[(C4-II)] $\rmE  \|\nabla\psi_{\theta_0}(Z)\|^4_2<\infty$;

\item[(C4-III)] $\rmE  \|\nabla^{\kappa}\psi_{\theta_0}(Z)\|^4_2<\infty,$  for $\kappa=2,3$. 

\end{enumerate}

}

All of the above conditions are mild and sensible, which are commonly used to guarantee the asymptotic normality and higher-order expansion of the {\color{black}estimator which solves the estimating equations (\ref{eq:estimating})} (see, e.g., \citealp{rilstone1996second,van2000asymptotic,kim2016higher}).

%\subsection{Proof of {\color{black}Lemma \ref{pn:1}}}
%\label{sec:proposition}

\subsection{Technical Lemmas}
\label{sec:lemma}

To facilitate the theoretical proofs, we provide {\color{black}five} important lemmas. The extensive proofs of these {\color{black}five} lemmas and {\color{black}13 additional} technical lemmas  can be found in the supplementary material.

 Let $Z_{1}, Z_{2}, \cdots \stackrel{iid}{\sim} Z$ and  $
 h(Z_{s_1},\cdots,Z_{s_{k_N}})\in\mathbb{R}$ 
 be any generic kernel function. Let
 \[
U_{k_N,m_N}=\frac{1}{m_N}\sum_{s\in\mathcal{S}}h(Z_{s_1},\cdots,Z_{s_{k_N}})
 \]
 be an incomplete, infinite order $U$-statistic with $0<k_N<N$ and $k_N\to\infty$. Let \be\label{eq:zeta}\zeta_{c,k_N}=\Cov\big(h(Z_1,\cdots,Z_c,Z_{c+1},\cdots,Z_{k_N}),h(Z_1,\cdots,Z_c,Z_{c+1}',\cdots,Z_{k_N}')\big)\ee for $1\leq c\leq k_N$, where $Z_{c+1}',\cdots,Z_{k_N}'$ $\stackrel{iid}\sim Z$ and are independent of $Z_1,Z_2,\cdots$.   For convenience, we denote $\rmE h^{\kappa}\triangleq \rmE h^{\kappa}(Z_1,\cdots,Z_{k_N})$ for $\kappa\geq 1$. It is worth noting that the asymptotic normality of $U_{k_N,m_M}$ in Lemma \ref{tm:0} below requires the following Lindeberg-type condition
 \be\label{eq:Lindeberg}
\rmE\lmk \lbk \frac{h_{1,k_N} (Z_1)}{\sqrt{\zeta_{1,k_N}}}\rbk^2I_{\lbk \lak\frac{h_{1,k_N} (Z_1)}{\sqrt{\zeta_{1,k_N}}}\rak>\epsilon\sqrt{N}\rbk}\rmk\to 0,
\ee
for all $\epsilon>0$ and $h_{1,k_N}(z_1)=\rmE\{ h(z_1,Z_2,\cdots,Z_{k_N}) \}-\rmE h$, where $I_{\{\cdot\}}$ is an indicator function.

  \begin{lemma}\label{tm:0} Assume $\rmE h^2<\infty$ given each $k_N$. We obtain
\[
\Var\lsk  U_{k_N,{m_N}}\rsk \leq \frac{k_N}{N} \zeta_{k_N,k_N}\lsk 1+ \frac{k_N}{N} \rsk+\frac{1}{m_N} \zeta_{k_N,k_N}\lbk 1-\frac{1}{{N\choose k_N}}\rbk.
\]
If we further assume   $\zeta_{1,k_N}\neq 0$, we have
\[
\Var\lsk  U_{k_N,{m_N}}\rsk=\frac{k_N^2}{N}\zeta_{1,k_N} \lbk 1+O\lsk a_N\rsk\rbk+\frac{1}{m_N} \zeta_{k_N,k_N}\lbk 1-\frac{1}{{N\choose k_N}}\rbk,
\]
where $a_N\triangleq (k_N/N)\{\zeta_{k_N,k_N}/(k_N\zeta_{1,k_N})\}$. In addition,  assume $k_N\to\infty$ and $a_N\to 0$ as $N\to\infty$. We then obtain

(1) If  $a_NN^2/(k_N^2m_N)\to 0$, then $\Var( U_{k_N,{m_N}})=\{1+o(1)\}k_N^2\zeta_{1,k_N}/N$. If we further assume the Lindeberg condition (\ref{eq:Lindeberg}) holds, then we have 
\[
\frac{ U _{k_N,m_N}-\rmE h}{\sqrt{k_N^2\zeta_{1,k_N}/N}}\stackrel{d}\longrightarrow \mN(0,1).
\]

(2)  If  $a_NN^2/(k_N^2m_N)\to \infty$,   then  $\Var( U_{k_N,{m_N}})= \{1+o(1)\}\zeta_{k_N,k_N}/m_N$. If we further assume, $m_N\to\infty$, $\rmE h^4<\infty$,  and  
$\rmE (h-\rmE h)^4/\{\rmE (h-\rmE h)^2\}^2=O(1)$, then we have 
\[
\frac{ U _{k_N,m_N}-\rmE h}{\sqrt{\zeta_{k_N,k_N}/m_N}}\stackrel{d}\longrightarrow \mN(0,1).
\]

(3) If  $a_NN^2/(k_N^2m_N)\to 1/\alpha\in (0,\infty)$,   then $\Var( U_{k_N,{m_N}})=\{1+o(1)\}(1+1/\alpha) k_N^2\zeta_{1,k_N}/N$. If we further assume the Lindeberg condition (\ref{eq:Lindeberg}) holds, $m_N\to\infty$, $\rmE h^4<\infty$,  and  
$\rmE (h-\rmE h)^4/\{\rmE (h-\rmE h)^2\}^2=O(1)$, then we have 
\[
\frac{ U _{k_N,m_N}-\rmE h}{\sqrt{(1+1/\alpha)k_N^2\zeta_{1,k_N}/N}}\stackrel{d}\longrightarrow \mN(0,1).
\]

\end{lemma}

 {\color{black}
 
 Let $h_{\theta_0}(Z_{s_1},\cdots,Z_{s_{k_N}})=\sum_{i=1}^{k_N}\psi_{\theta_0}(Z_{s_i})$, and 
 \be\label{eq:}
\mU_{\theta_0,k_N,m_N}= \frac{1}{m_N}\sum_{s\in\mathcal{S}}h_{\theta_0}(Z_{s_1},\cdots,Z_{s_{k_N}}).
\ee
We then obtain the following lemma.

 \begin{lemma}\label{tm:1}Assume $\rmE \|\psi_{\theta_0}(Z)\|_2^2<\infty$, and  $\Sigma_{\theta_0}$ is not singular. We have
\[
\Var\lsk  \mU_{\theta_0,k_N,{m_N}}\rsk=\frac{k_N^2}{N}\Sigma_{\theta_0} \lbk 1+O\lsk \frac{k_N}{N}\rsk\rbk+\frac{k_N}{m_N} \Sigma_{\theta_0}\lbk 1-\frac{1}{{N\choose k_N}}\rbk.
\]
In addition,  assume $k_N\to\infty$ and $k_N/N\to 0$ as $N\to\infty$. We then obtain

(1) If  $(k_Nm_N)/N\to \infty$, then $\Var( \mU_{\theta_0,k_N,{m_N}})=\{1+o(1)\}k_N^2\Sigma_{\theta_0}/N$, and
\[
\frac{  \mU_{\theta_0,k_N,{m_N}}}{\sqrt{k_N^2/N}}\stackrel{d}\longrightarrow \mN(0,\Sigma_{\theta_0}).
\]

(2)  If  $(k_Nm_N)/N\to 0$,   then  $\Var( \mU_{\theta_0,k_N,{m_N}})= \{1+o(1)\}k_N\Sigma_{\theta_0}/m_N$. If we further assume $m_N\to\infty$, and $\rmE \|\psi_{\theta_0}(Z)\|_2^4<\infty$, then we have 
\[
\frac{  \mU_{\theta_0,k_N,{m_N}}}{\sqrt{k_N/m_N}}\stackrel{d}\longrightarrow \mN(0,\Sigma_{\theta_0}).
\]

(3) If  $(k_Nm_N)/N\to \alpha \in (0,\infty)$,   then $\Var( \mU_{\theta_0,k_N,{m_N}})=\{1+o(1)\}(1+1/\alpha) k_N^2\Sigma_{\theta_0}/N$. If we further assume  $\rmE \|\psi_{\theta_0}(Z)\|_2^4<\infty$, then we have 
\[
\frac{ \mU_{\theta_0,k_N,{m_N}}}{\sqrt{(1+1/\alpha)k_N^2/N}}\stackrel{d}\longrightarrow \mN(0,\Sigma_{\theta_0}).
\]

 \end{lemma}

%It is worth noting that Theorem 5.21 in \citet[p. 52]{van2000asymptotic} has already given the asymptotic normality of the estimator which solves the estimating equations (\ref{eq:estimating}). 

%However, in order to prepare for our further analysis, we combine it with the higher-order expansion of the {\color{black}estimator} (see, e.g., \citealp{rilstone1996second,kim2016higher}), and explicitly show the higher-order bias in the following lemma.  

 Following the notations defined before Theorem \ref{tm:2}, we obtain the following lemma for the subsample estimator $\hat\theta_{k_N,s}$ which solves the estimating equations (\ref{eq:estimating}).

\begin{lemma} \label{pn:1}Under Conditions (C1), (C2-I) -- (C2-II) and (C3-I) -- (C3-II) in Appendix \ref{sec:c}, for $0<k_N\leq N$ and $k_N\to\infty$,  we obtain   
\[
\sqrt{k_N}(\hat\theta_{k_N,s}-\theta_0)=-V_{\theta_0}^{-1}J_{k_N,s}+\frac{\mathcal{B}_{k_N,s}}{\sqrt{k_N}}+O_{\rmP}\lsk \frac{1}{k_N}\rsk\stackrel{d}\longrightarrow \mN\Big(0,V_{\theta_0}^{-1} \Sigma_{\theta_0}\lsk V_{\theta_0}^{-1}\rsk^\top\Big) 
\]
uniformly for $s\in S_{N}^{k_N}$, where $
\rmE(\mathcal{B}_{k_N,s})=-V_{\theta_0}^{-1}\{ -H_{\theta_0}\rmvec( V_{\theta_0}^{-1}) +0.5V^{(2)}_{\theta_0}( V_{\theta_0}^{-1}\otimes V_{\theta_0}^{-1}) \rmvec( \Sigma_{\theta_0})\}$ $=B_{\theta_0},
$
 and $\mathcal{B}_{k_N,s}=O_{\rmP}(1)$ uniformly for $s\in S_{N}^{k_N}$. 
\end{lemma}

\begin{lemma} \label{pn:2} Let $\hat\theta_{k_N,s}^{(bc)}$ be any one of the three bias-corrected subsample estimators in Section \ref{sec:sbc}. 
Under Conditions (C1) -- (C4) in Appendix \ref{sec:c}, for $0<k_N\leq N$ and $k_N\to\infty$,  we obtain 
\[
\sqrt{{k_N}}(\hat\theta_{k_N,s}^{(bc)}-\theta_0)=-V_{\theta_0}^{-1}J_{k_N,s}+\frac{\mathcal{B}_{k_N,s}-B_{\theta_0}}{\sqrt{{k_N}}}+O_{\rmP}\lsk \frac{1}{{k_N}}\rsk\stackrel{d}\longrightarrow \mN\Big(0,V_{\theta_0}^{-1} \Sigma_{\theta_0}\lsk V_{\theta_0}^{-1}\rsk^\top\Big),
\]
uniformly for $s\in S_{N}^{k_N}$.

\end{lemma}

\begin{lemma}\label{la:bs} Under Conditions (C4-I) -- (C4-II) in Appendix \ref{sec:c},  for  $\mathcal{B}_{k_N,s}$ defined in (\ref{eq:bkn}), we obtain
\[
\Var\lsk \frac{1}{m_N}\sum_{s\in\mS} k_N \mathcal{B}_{k_N,s}\rsk=O\lsk  \frac{k_N^3}{N} \lsk 1+ \frac{k_N}{N} \rsk+\frac{k_N^2}{m_N}\lbk 1-\frac{1}{{N\choose k_N}}\rbk\rsk.
\]
\end{lemma}

}

\subsection{Proofs of Theorems}

\label{sec:proofs} 

{\color{black}

\noindent\textbf{Proof of Theorem \ref{tm:0}}.  We employ Lemma \ref{tm:1} by defining 
 $\psi_{\theta_0}(Z_i)=Z_i-\mu_0$ based on Example \ref{ex:1}. We then obtain all the results  for $\mU_{\theta_0,k_N,m_N}$ in Lemma \ref{tm:1}. Note that $\mu_{k_N,m_N}-\mu_0=k_N^{-1}\mU_{\theta_0,k_N,m_N}$. This, together with Lemma \ref{tm:1}, leads to the desired result.

\bigskip

}

\noindent\textbf{Proof of Theorem \ref{tm:2}}. 
{\color{black}Since Lemma  \ref{pn:1} holds uniformly for $s\in S_{N}^{k_N}$}, we aggregate $m_N$ subsample {\color{black}estimator}s by (\ref{eq:emeq}) and obtain {\color{black}the asymptotic expansion}
\be\label{eq:expansion}
\hat\theta_{k_N,m_N}-\theta_0=-\frac{1}{k_N}V_{\theta_0}^{-1}\mU_{\theta_0,k_N,m_N}+\frac{1}{m_N}\sum_{s\in\mS}\frac{\mathcal{B}_{k_N,s}}{{k_N}}+O_{\rmP}\lsk \frac{1}{{k_N^{3/2}}}\rsk,
\ee
{\color{black}where 
$\mU_{\theta_0,k_N,m_N}$ is defined in (\ref{eq:})}. {\color{black}Then the expansion in the theorem can also be obtained since $m_N^{-1}\sum_{s\in\mS}J_{k_N,s}=k_N^{-1/2}\mU_{\theta_0,k_N,m_N}$.}

{\color{black}We subsequently obtain the results for $
\rmE(m_N^{-1}\sum_{s\in\mS}{\mathcal{B}_{k_N,s}})=-V_{\theta_0}^{-1}\{ -H_{\theta_0}\rmvec( V_{\theta_0}^{-1}) +0.5V^{(2)}_{\theta_0}( V_{\theta_0}^{-1}\otimes V_{\theta_0}^{-1}) \rmvec( \Sigma_{\theta_0})\}= B_{\theta_0} 
$  
 and $m_N^{-1}\sum_{s\in\mS}{\mathcal{B}_{k_N,s}}=O_{\rmP}(1)$ by Lemma \ref{pn:1}. Using the latter result, along with Lemma \ref{tm:1}, we show the results in this theorem under settings (i) -- (ii), respectively.

 }
%$\rmE (k_N^{-1}V_{\theta_0}^{-1}\mU_{\theta_0,k_N,m_N})=0$, and
%\[
%\Var\lsk -\frac{1}{k_N}V_{\theta_0}^{-1}\mU_{\theta_0,k_N,m_N} \rsk=  V_{\theta_0}^{-1} \Sigma_{\theta_0}\lsk V_{\theta_0}^{-1}\rsk^\top\lmk\frac{1}{N} \lbk 1+O\lsk \frac{k_N}{N}\rsk\rbk+\frac{1}{m_Nk_N} \lbk 1-\frac{1}{{N\choose k_N}}\rbk\rmk
%\]
%by {\color{black}Lemma \ref{tm:1}}. 

%Using Lemma \ref{la:bs}, we also obtain
%\be\label{eq:bias}
%\rmE\lsk \frac{1}{m_N}\sum_{s\in\mS}\frac{\mathcal{B}_{k_N,s}}{{k_N}}\rsk=\frac{B_{\theta_0}}{k_N}=O\lsk \frac{1}{k_N}\rsk,\textrm{ and }
%\ee
%\be\label{eq:varaince}
%\Var\lsk \frac{1}{m_N}\sum_{s\in\mS}\frac{\mathcal{B}_{k_N,s}}{{k_N}}\rsk=O\lsk  \frac{1}{k_NN} \lsk 1+ \frac{k_N}{N} \rsk+\frac{1}{k_Nm_Nk_N}\lbk 1-\frac{1}{{N\choose k_N}}\rbk\rsk.
%\ee
%These conclusions lead to the bias and variance of $\hat\theta_{k_N,m_N}$ in this theorem as $N\to\infty$ and $k_N\to \infty$. 

{\color{black}
Under setting (i), we obtain $m_N^{-1}\sum_{s\in\mS}$ ${\mathcal{B}_{k_N,s}}/{{k_N}}=O_{\rmP}(k_N^{-1})$.  We then apply the expansion (\ref{eq:expansion}) and  scenarios (1) and (3) of {\color{black}Lemma \ref{tm:1}}, and hence obtain the asymptotic normality result under  $k_N/\sqrt{N}\to \infty$ by Slutsky's theorem.}

{\color{black}
Under setting (ii), we also have $m_N^{-1}\sum_{s\in\mS}{\mathcal{B}_{k_N,s}}/{{k_N}}=O_{\rmP}(k_N^{-1})$.  We then apply the expansion (\ref{eq:expansion}) and scenario (2) of {\color{black}Lemma \ref{tm:1}}, and hence obtain the asymptotic normality result under  $k_N/m_N\to \infty$ by Slutsky's theorem, which completes the entire proof.

}

\bigskip

{\color{black}
\noindent\textbf{Proof of Theorem \ref{cy:2}}.  
First, we re-express $\hat\Sigma^{(bc)}_{k_N,s}$ in Example \ref{ex:3} by
\[
\hat\Sigma^{(bc)}_{k_N,s}
=\frac{1}{k_N-1}\sum_{i=1}^{k_N} (Z_{s_i}-\mu_0)(Z_{s_i}-\mu_0)^\top-\frac{1}{k_N(k_N-1)} \lbk  \sum_{i=1}^{k_N} (Z_{s_i}-\mu_0)\rbk\lbk \sum_{i=1}^{k_N} (Z_{s_i}-\mu_0)\rbk^\top.
\]
Then, we can partition $\hat\theta^{(bc)}_{k_N,m_N}$ into two parts by
 \[
\hat\theta^{(bc)}_{k_N,m_N}= \mU_N^{(1)}+\mU_N^{(2)}
\]
where
\[
\mU_N^{(1)}=\frac{1}{m_N}\sum_{s\in\mS} \begin{pmatrix} \frac{1}{k_N}\sum_{i=1}^{k_N} Z_{s_i}\\L_p\rmvec\lmk 
\frac{1}{k_N-1}\sum_{i=1}^{k_N} (Z_{s_i}-\mu_0)(Z_{s_i}-\mu_0)^\top\rmk
\end{pmatrix}\textrm{ and }
\]
\[
\mU_N^{(2)}=-\frac{1}{m_N}\sum_{s\in\mS} \begin{pmatrix} 0_{p\times 1}\\L_p\rmvec\lmk 
\frac{1}{k_N(k_N-1)} \lbk  \sum_{i=1}^{k_N} (Z_{s_i}-\mu_0)\rbk\lbk \sum_{i=1}^{k_N} (Z_{s_i}-\mu_0)\rbk^\top\rmk
\end{pmatrix},
\]
with $0_{p_1\times p_2}$ denoting a $p_1\times p_2$ matrix of zeros.

To facilitate the proof, let
\[
\tilde \theta^{(bc)}_{k_N,m_N}=\begin{pmatrix} I_p & 0_{p\times p(p+1)/2}\\
0_{p(p+1)/2\times p} & \frac{k_N-1}{k_N} I_{p(p+1)/2}
\end{pmatrix}\mU_N^{(1)},
\]
where $I_p$ is the $p$-dimensional identity matrix. 
We then have
\be\label{eq:t13}
\hat\theta^{(bc)}_{k_N,m_N}-\theta_0= \begin{pmatrix} I_p & 0\\
0 & \frac{k_N}{k_N-1} I_{p(p+1)/2}
\end{pmatrix}\lsk \tilde \theta^{(bc)}_{k_N,m_N}-\theta_0\rsk+ 
\begin{pmatrix}
0_{p\times 1}\\
\frac{1}{k_N-1}\vech (\Sigma_0)
\end{pmatrix}  +\mU_N^{(2)}.
\ee

It is worth noting that $\tilde \theta^{(bc)}_{k_N,m_N}$ in (\ref{eq:t13}) is an incomplete $U$-statistic. We employ Lemma \ref{tm:1} by defining 
 $\psi_{\theta_0}(Z_i)=\begin{pmatrix}Z_i-\mu_0\\L_p\rmvec\{ 
 (Z_{s_i}-\mu_0)(Z_{s_i}-\mu_0)^\top-\Sigma_0\}\end{pmatrix}$. This directly leads to that  that under the conditions given in Corollary \ref{cy:1}, 
 all the results in Corollary \ref{cy:1} hold for $\tilde\theta_{k_N,m_N}^{(bc)}$ without requiring $k_N/\sqrt{N}\to \infty$ in (i) and  $k_N/m_N\to \infty$ in (ii).

Next, we focus on the last two terms on the right hand side of (\ref{eq:t13}), the sum of which is equal to
\bda
-\frac{1}{m_N}\sum_{s\in\mS} \begin{pmatrix} 0_{p\times 1}\\L_p\rmvec\lmk 
\frac{1}{k_N(k_N-1)} \lbk  \sum_{i=1}^{k_N} (Z_{s_i}-\mu_0)\rbk\lbk \sum_{i=1}^{k_N} (Z_{s_i}-\mu_0)\rbk^\top-\frac{1}{k_N-1}\Sigma_0\rmk
\end{pmatrix}.
\eda
 Consider arbitrary $t^{(1)},t^{(2)}\in\mathbb{R}^p$ and define $U_N^{(2)}=$
\[
\frac{1}{m_N}\sum_{s\in\mS}\lmk \frac{1}{k_N(k_N-1)} t^{(1)\top} \lbk  \sum_{i=1}^{k_N} (Z_{s_i}-\mu_0)\rbk\lbk \sum_{i=1}^{k_N} (Z_{s_i}-\mu_0) \rbk^\top  t^{(2)}- \frac{1}{k_N-1}t^{(1)\top}\Sigma t^{(2)}\rmk.
\]

Using (\ref{eq:t13}), the asymptotic results for $\tilde \theta^{(bc)}_{k_N,m_N}$, and 
Slutsky's theorem, 
to prove Theorem \ref{cy:2} it suffices to show that
\[
\frac{ 1}{\sqrt{(1+1/\alpha)/N}} U_N^{(2)}=o_{\rmP}(1)
\]
under setting (i) and 
\[
\frac{ 1}{\sqrt{1/(k_Nm_N)}} U_N^{(2)}=o_{\rmP}(1)
\]
under setting (ii), due to the arbitrary $t^{(1)},t^{(2)}\in\mathbb{R}^p$.
  As $\rmE \,  U_N^{(2)}=0$, we only need to derive the order of $\Var (U_N^{(2)})$. 

To facilitate the proof, consider a kernel function 
\[
\tilde {\tilde h}(Z_{s_1},\cdots,Z_{s_{k_N}})=\frac{1}{k_N(k_N-1)} t^{(1)\top} \lbk  \sum_{i=1}^{k_N} (Z_{s_i}-\mu_0)\rbk\lbk \sum_{i=1}^{k_N} (Z_{s_i}-\mu_0) \rbk^\top  t^{(2)}.
\]
Then $
\rmE \tilde {\tilde h} = \frac{1}{k_N-1}t^{(1)\top}\Sigma t^{(2)}$. 
Due to the expression of $\tilde {\tilde h}$ and using similar techniques in the
proof of (\ref{eq:eh4}) in the supplementary material, we obtain 
\[
\rmE \tilde {\tilde h}^2=\frac{1}{k_N^2(k_N-1)^2} O(k_N^2)=O\lsk \frac{1}{k_N^2}\rsk.
\]
Consequently, we obtain $\Var ( \tilde {\tilde h})=\rmE \tilde {\tilde h}^2-(\rmE \tilde {\tilde h})^2=O(k_N^{-2})$.
Using Lemma \ref{tm:0} and the fact that $U_N^{(2)}+\rmE \tilde {\tilde h}$ is an incomplete $U$-statistics with kernel $\tilde {\tilde h}$, we have
\[
\Var (U_N^{(2)})=O\lsk  \frac{1}{N k_N} \rsk+O\lsk \frac{1}{m_Nk_N^2}\rsk.
\]
Under setting (i), if $k_Nm_N/N\to \alpha\in(0,\infty]$, we get
\ben\label{eq:t14}
\Var (U_N^{(2)})=O\lsk  \frac{1}{N k_N} \rsk\Rightarrow  \frac{ 1}{\sqrt{(1+1/\alpha)/N}} U_N^{(2)}=O_{\rmP}\lsk \frac{1}{\sqrt{k_N}}\rsk=o_{\rmP}(1).
\een
Under setting (ii), if $k_Nm_N/N\to 0$, we obtain
\ben\label{eq:t15}
\Var (U_N^{(2)})=O\lsk \frac{1}{m_Nk_N^2} \rsk\Rightarrow  \frac{ 1}{\sqrt{1/(k_Nm_N)}}U_N^{(2)}=O_{\rmP}\lsk \frac{1}{\sqrt{k_N}}\rsk=o_{\rmP}(1),
\een
which accomplishes the entire proof.

}

\bigskip

\noindent\textbf{Proof of Theorem \ref{tm:3}}. 
{\color{black}Lemma \ref{pn:2}} provides the higher-order expansion of the bias-corrected  subsample {\color{black}estimator} $\hat\theta_{k_N,s}^{(bc)}$ {\color{black}uniformly for $s\in S_{N}^{k_N}$}. Then we aggregate $m_N$ bias-corrected subsample {\color{black}estimator}s by {\color{black}the equation above Theorem \ref{tm:3}} and obtain
\be\label{eq:expansion1}
\hat\theta_{k_N,m_N}^{(bc)}-\theta_0=-\frac{1}{k_N}V_{\theta_0}^{-1}\mU_{\theta_0,k_N,m_N}+\widetilde \mU_{\theta_0,k_N,m_N} +O_{\rmP}\lsk \frac{1}{{k_N^{3/2}}}\rsk,
\ee
where $\mU_{\theta_0,k_N,m_N}$ is defined in (\ref{eq:}), {\color{black}and $\widetilde \mU_{\theta_0,k_N,m_N}=m_N^{-1}\sum_{s\in\mS}{(\mathcal{B}_{k_N,s}-B_{\theta_0})}/{{k_N}}$.}
{\color{black}By Theorem \ref{tm:2} and Lemma \ref{la:bs}, we have}
\be\label{eq:bias1}
\rmE\lsk \widetilde \mU_{\theta_0,k_N,m_N}\rsk=0,\textrm{ and }
\ee
\be\label{eq:varaince1}
\Var\lsk \widetilde \mU_{\theta_0,k_N,m_N}\rsk=O\lsk  \frac{1}{k_NN} \lsk 1+ \frac{k_N}{N} \rsk+\frac{1}{k_Nm_Nk_N}\lbk 1-\frac{1}{{N\choose k_N}}\rbk\rsk.
\ee
{\color{black}Next} we show the results in this theorem under  settings (i) -- (ii), respectively.

Under setting (i), {\color{black}we have}  $\Var( \widetilde \mU_{\theta_0,k_N,m_N} )=$ $O(k_N^{-1}N^{-1})$ by (\ref{eq:varaince1}). This result, together with  (\ref{eq:bias1}), leads to $ \widetilde \mU_{\theta_0,k_N,m_N}=O_{\rmP}(k_N^{-1/2}N^{-1/2})=O_{\rmP}(k_N^{-3/2})$ since $k_N=O(\sqrt{N})$.  
We then apply the expansion (\ref{eq:expansion1}) and {\color{black}Lemma \ref{tm:1}}, and hence obtain the asymptotic normality result under  $k_N/\sqrt[3]{N}\to \infty$ by Slutsky's theorem.

Under setting (ii),  {\color{black}we have} $\Var(  \widetilde \mU_{\theta_0,k_N,m_N})=O(k_N^{-1}m_N^{-1}k_N^{-1})$ by (\ref{eq:varaince1}). This result, together with (\ref{eq:bias1}), leads to $\color{black} \widetilde \mU_{\theta_0,k_N,m_N}=O_{\rmP}(k_N^{-1/2}m_N^{-1/2}k_N^{-1/2})=O_{\rmP}(k_N^{-3/2})$ since $k_N=O(m_N)$.  
We then apply the expansion (\ref{eq:expansion1}) and {\color{black}Lemma \ref{tm:1}}, and hence obtain the asymptotic normality result under  $k_N/\sqrt{m_N}\to \infty$ by Slutsky's theorem, which completes the entire proof.

\bigskip

\noindent\textbf{Proof of Theorem \ref{tm:4}}.  The subbagging variance estimators can be re-expressed by
\[
\widehat\Omega_{k_N,m_N}=\frac{1}{m_N}\sum_{s\in\mS}\lsk \hat\theta_{k_N,s}-\theta_0\rsk \lsk \hat\theta_{k_N,s}- \theta_0\rsk^\top-\lsk  \hat\theta_{k_N,m_N}-\theta_0\rsk \lsk  \hat\theta_{k_N,m_N}-\theta_0\rsk^\top,\textrm{ and }
\]
\[ \widehat\Omega^{(bc)}_{k_N,m_N}=\frac{1}{m_N}\sum_{s\in\mS}\lsk \hat\theta^{(bc)}_{k_N,s}-\theta_0\rsk \lsk \hat\theta_{k_N,s}^{(bc)}- \theta_0\rsk^\top-\lsk \hat\theta^{(bc)}_{k_N,m_N}-\theta_0\rsk\lsk \hat\theta^{(bc)}_{k_N,m_N}-\theta_0\rsk^\top.
\]

We first show the consistency of $\widehat\Omega_{k_N,m_N}$.  Under the conditions given in Theorem \ref{tm:2}, we obtain
\[
\frac{1}{m_N}\sum_{s\in\mS}\lsk \hat\theta_{k_N,s}-\theta_0\rsk \lsk \hat\theta_{k_N,s}- \theta_0\rsk^\top=\frac{1}{m_N}\sum_{s\in\mS} V_{\theta_0}^{-1}\frac{J_{k_N,s}J_{k_N,s}^\top}{k_N}\lsk V_{\theta_0}^{-1}\rsk^\top+O_{\rmP}\lsk \frac{1}{k_N^{3/2}}\rsk
\]
by {\color{black}Lemma \ref{pn:1}}. Consequently,
\be\label{eq:hatomega}
k_N \widehat\Omega_{k_N,m_N}=\frac{1}{k_N}\widetilde{\widetilde \mU}_{\theta_0,k_N,m_N}+O_{\rmP}\lsk \frac{1}{k_N^{1/2}}\rsk +O_{\rmP}\lsk k_N\laak  \hat\theta_{k_N,m_N}-\theta_0 \raak_2^2\rsk,
\ee
where the leading term 
\[
\widetilde{\widetilde \mU}_{\theta_0,k_N,m_N}=\frac{1}{m_N}\sum_{s\in\mS} V_{\theta_0}^{-1}{\frac{J_{k_N,s}}{k_N^{-1/2}}\frac{J_{k_N,s}^\top}{k_N^{-1/2}}}\lsk V_{\theta_0}^{-1}\rsk^\top
\]
is an incomplete $U$-statistic with the kernel being $V_{\theta_0}^{-1}{{J_{k_N,s}}{k_N^{1/2}}{J_{k_N,s}^\top}{k_N^{1/2}}}( V_{\theta_0}^{-1})^\top$, and its mean is
\be\label{eq:meanu}
\rmE\lmk V_{\theta_0}^{-1}\lbk \sum_{i=1}^{k_N}\psi_{\theta_0}(Z_{s_i})\rbk\lbk \sum_{i=1}^{k_N}\psi_{\theta_0}(Z_{s_i})\rbk^\top\lsk V_{\theta_0}^{-1}\rsk^\top \rmk=k_N V_{\theta_0}^{-1}\Sigma_{\theta_0}\lsk V_{\theta_0}^{-1}\rsk^\top.
\ee
Next we apply Lemma \ref{tm:0} to derive the variance rate of $\widetilde{\widetilde \mU}_{\theta_0,k_N,m_N}$.  Consider arbitrary $t^{(1)},t^{(2)}\in\mathbb{R}^d$. Then it is sufficient to derive the variance rate of $m_N^{-1}\sum_{s\in\mS}  \tilde {\tilde h}_{\theta_0}(Z_{s_1},\cdots,Z_{s_{k_N}})$, where $ \tilde {\tilde h}_{\theta_0}(Z_{s_1},\cdots,Z_{s_{k_N}})=t^{(1)}{}^\top V_{\theta_0}^{-1}{{J_{k_N,s}}{k_N^{1/2}}{J_{k_N,s}^\top}{k_N^{1/2}}}( V_{\theta_0}^{-1})^\top t^{(2)}$. In fact, we know this kernel function can be re-expressed by $
 \tilde {\tilde h}_{\theta_0}= t^{(1)}{}^\top V_{\theta_0}^{-1} h_{\theta_0} h_{\theta_0}^\top( V_{\theta_0}^{-1})^\top t^{(2)}$, which 
is actually from the operation of kernel $h_{\theta_0}$ {\color{black}defined before (\ref{eq:})}. In order to apply Lemma \ref{tm:0}, we require $\rmE \tilde {\tilde h}_{\theta_0}^2<\infty$ given each $k_N$. As $|\tilde {\tilde h}_{\theta_0}|\leq \|t^{(1)}\|_2 \|t^{(2)}\|_2 \|V_{\theta_0}^{-1}\|_2^2 \|h_{\theta_0}\|_2^2$, we get $\rmE \tilde {\tilde h}_{\theta_0}^2\leq \|t^{(1)}\|_2^2 \|t^{(2)}\|_2^2 \|V_{\theta_0}^{-1}\|_2^4 \rmE\|h_{\theta_0}\|_2^4$. Thus it is sufficient to require $\rmE \|h_{\theta_0}\|_2^4<\infty$, which is  obtained by Condition {\color{black}(C4-I)}. Hence, we can use the inequality in Lemma \ref{tm:0} to give the order of the variance. We define $\tilde{\tilde \zeta}_{\theta_0, c,k_N}$ by (\ref{eq:zeta}) via replacing the kernel $h$ by $\tilde{\tilde h}_{\theta_0}$. Using Lemma \ref{tm:0}, we then obtain
\be\label{eq:varth1}
\Var\lsk \frac{1}{m_N}\sum_{s\in\mS} \tilde {\tilde h}_{\theta_0}(Z_{s_1},\cdots,Z_{s_{k_N}})\rsk \leq \frac{k_N}{N} \tilde{\tilde \zeta}_{\theta_0,k_N,k_N}\lsk 1+ \frac{k_N}{N} \rsk+\frac{1}{m_N}\tilde{\tilde \zeta}_{\theta_0,k_N,k_N}\lbk 1-\frac{1}{{N\choose k_N}}\rbk.
\ee
Now the key is to derive $ \tilde{\tilde \zeta}_{\theta_0,k_N,k_N}=\Var(\tilde{\tilde h}_{\theta_0})=\rmE(\tilde{\tilde h}_{\theta_0}^2)-(\rmE\tilde{\tilde h}_{\theta_0})^2$. First, by (\ref{eq:meanu}) we have, 
\be\label{eq:Eth01}
\rmE\tilde{\tilde h}_{\theta_0}=k_N t^{(1)}{}^\top V_{\theta_0}^{-1}\Sigma_{\theta_0}\lsk V_{\theta_0}^{-1}\rsk^\top t^{(2)}=O(k_N).
\ee
Second, 
\[
\rmE(\tilde {\tilde h}_{\theta_0}^2)=\rmE\lbk   t^{(1)}{}^\top V_{\theta_0}^{-1} h_{\theta_0} h_{\theta_0}^\top( V_{\theta_0}^{-1})^\top t^{(2)} t^{(2)}{}^\top V_{\theta_0}^{-1} h_{\theta_0} h_{\theta_0}^\top( V_{\theta_0}^{-1})^\top t^{(1)}\rbk.
\]
Due to this expression and using similar techniques  in the 
proof of (\ref{eq:eh4}), we obtain $\rmE(\tilde {\tilde h}_{\theta_0}^2)=O(k_N^2)$. This, together with (\ref{eq:Eth01}), leads to $ \tilde {\tilde \zeta}_{\theta_0,k_N,k_N}=\Var(\tilde {\tilde h}_{\theta_0})=\rmE(\tilde {\tilde h}_{\theta_0}^2)-(\rmE\tilde {\tilde h}_{\theta_0})^2=O(k_N^2)$. Consequently, based on (\ref{eq:varth1}) we obtain
\ben
\Var\lsk \frac{1}{m_N}\sum_{s\in\mS} \tilde {\tilde h}_{\theta_0}(Z_{s_1},\cdots,Z_{s_{k_N}})\rsk =O\lsk  \frac{k_N^3}{N} \lsk 1+ \frac{k_N}{N} \rsk+\frac{k_N^2}{m_N}\lbk 1-\frac{1}{{N\choose k_N}}\rbk\rsk,
\een
which leads to
\be\label{eq:varo}
\Var\lsk\frac{1}{k_N} \widetilde{\widetilde \mU}_{\theta_0,k_N,m_N}\rsk =O\lsk  \frac{k_N}{N} \lsk 1+ \frac{k_N}{N} \rsk+\frac{1}{m_N}\lbk 1-\frac{1}{{N\choose k_N}}\rbk\rsk=o(1)
\ee
 due to the arbitrary $t^{(1)},t^{(2)}\in\mathbb{R}^d$, $k_N/N\to 0$ and $m_N\to\infty$.

Under setting (i) in Theorem \ref{tm:2}, $\color{black}\|  \hat\theta_{k_N,m_N}-\theta_0 \|_2^2=O_{\rmP}(N^{-1})$. This result, in conjunction with (\ref{eq:hatomega}) and (\ref{eq:varo}), leads to $k_N \widehat\Omega_{k_N,m_N}-V_{\theta_0}^{-1}\Sigma_{\theta_0}( V_{\theta_0}^{-1})^\top=o_{\rmP}(1)+O_{\rmP}(k_N^{-1/2})+O_{\rmP}(k_N/N)=o_{\rmP}(1)$ by Chebyshev's inequality.

Under setting (ii) in Theorem \ref{tm:2}, $\color{black}\|  \hat\theta_{k_N,m_N}-\theta_0 \|_2^2=O_{\rmP}(k_N^{-1}m_N^{-1})$. This result, in conjunction with (\ref{eq:hatomega}) and (\ref{eq:varo}), leads to $k_N \widehat\Omega_{k_N,m_N}-V_{\theta_0}^{-1}\Sigma_{\theta_0}( V_{\theta_0}^{-1})^\top=o_{\rmP}(1)+O_{\rmP}(k_N^{-1/2})+O_{\rmP}(m_N^{-1})=o_{\rmP}(1)$ by Chebyshev's inequality, which completes the proof for $\widehat\Omega_{k_N,m_N}$.

We next show $\widehat\Omega^{(bc)}_{k_N,m_N}$.  Under the conditions given in Theorem \ref{tm:3}, we obtain
\[
\frac{1}{m_N}\sum_{s\in\mS}\lsk \hat\theta^{(bc)}_{k_N,s}-\theta_0\rsk \lsk \hat\theta^{(bc)}_{k_N,s}- \theta_0\rsk^\top=\frac{1}{m_N}\sum_{s\in\mS} V_{\theta_0}^{-1}\frac{J_{k_N,s}J_{k_N,s}^\top}{k_N}\lsk V_{\theta_0}^{-1}\rsk^\top+O_{\rmP}\lsk \frac{1}{k_N^{3/2}}\rsk
\]
by {\color{black}Lemma \ref{pn:2}}.  
Consequently,
\ben
k_N \widehat\Omega^{(bc)}_{k_N,m_N}=\frac{1}{k_N}\widetilde{\widetilde \mU}_{\theta_0,k_N,m_N}+O_{\rmP}\lsk \frac{1}{k_N^{1/2}}\rsk +O_{\rmP}\lsk k_N\laak  \hat\theta_{k_N,m_N}^{(bc)}-\theta_0 \raak_2^2\rsk,
\een
which has the same expansion as $k_N \widehat\Omega_{k_N,m_N}$. Using similar techniques  in the
proof of $k_N\widehat\Omega_{k_N,m_N}$, we obtain the  consistency of $k_N \widehat\Omega^{(bc)}_{k_N,m_N}$ under the conditions given in Theorem \ref{tm:3}, which completes the entire proof.

\bigskip

 \normalsize
 \spacingset{1.1}
\bibliographystyle{apalike}
 \bibliography{Reference}

 \normalsize
 \spacingset{1.45} 
 
\vspace{2cm}

 \begin{table}[htbp!]
\caption{The average bias (BIAS), standard deviation (SD), root mean
squared error (RMSE) and memory usage ({\color{black}MEMORY, in KiB}) of the full sample estimates. {Values of BIAS, SD and RMSE in this table are 100 times  their original values.}} \label{tb:1}

\begin{center}
\scalebox{1}{
\begin{tabular}{c|rr|rr|rr}
  \hline%full sample {\color{black}estimator}

\multicolumn{7}{c}{Full Sample Estimate {\color{black} for Logistic Regression}}   \\
  \hline
& \multicolumn{2}{c|}{\makecell[cb]{{$N$=2,000}}}    & \multicolumn{2}{c|}{\makecell[cb]{{$N$=10,000}}}      &
\multicolumn{2}{c}{\makecell[cb]{{$N$=50,000}}}   \\
\cline{2-3}\cline{4-5}\cline{6-7}
  \multicolumn{1}{c|}{}& \multicolumn{1}{c}{$\hat\theta_1$}& \multicolumn{1}{c|}{$\hat\theta_2$} &  \multicolumn{1}{c}{$\hat\theta_1$} & \multicolumn{1}{c|}{$\hat\theta_2$} &  \multicolumn{1}{c}{$\hat\theta_1$} & \multicolumn{1}{c}{$\hat\theta_2$}  \\
   \hline
  BIAS &-0.14     &0.42   &0.03      &-0.09  &-0.01     &-0.03           \\
 SD &4.92      &5.99  &2.22      &2.67 &1.00       &1.16    \\
  RMSE  &4.92      &6.01 &2.22      &2.67 &1.00      &1.16      \\
\multicolumn{1}{c|}{ MEMORY}  & \multicolumn{2}{c|}{48} &  \multicolumn{2}{c|}{240} &   \multicolumn{2}{c}{1200}   \\
\hline

\end{tabular}}
\end{center}

\end{table}

\begin{table}[htbp!]
\caption{The average bias (BIAS), standard deviation (SD), root mean
squared error (RMSE), asymptotic standard deviation (ASD),  average subbagging standard error (SSE), empirical coverage probability (CP), and memory usage ({\color{black}MEMORY, in KiB}) of the subbagging estimates. Values of BIAS, SD, RMSE, ASD and SSE in this table are 100 times  their original values. Only the settings in bold satisfy the requirement of Algorithm  \ref{al:1}.} \label{tb:2}

\begin{center}
\scalebox{0.66}{
\begin{tabular}{c|c|rr|rr|rr|rr||rr|rr}
   \hline%full sample {\color{black}estimator}

\multicolumn{14}{c}{Algorithm \ref{al:1} {\color{black} for Logistic Regression}}   \\
  \hline
   && \multicolumn{12}{c}{$\alpha=1$}\\  
\cline{3-4}\cline{5-6}\cline{7-8}\cline{9-10}\cline{11-12}\cline{13-14} 
 && \multicolumn{2}{c|}{$ k_N=\lfloor N^{5/12}\rfloor$}    & \multicolumn{2}{c|}{$k_N=\lfloor N^{6/12}\rfloor$}      &
\multicolumn{2}{c|}{$\boldsymbol {k_N=\lfloor N^{7/12}\rfloor}$}   & \multicolumn{2}{c||}{$\boldsymbol {k_N=\lfloor N^{8/12}\rfloor}$}&  \multicolumn{2}{c|}{$k_N=\lfloor N^{5/12}\rfloor$}
&  \multicolumn{2}{c}{$\boldsymbol {k_N=\lfloor N^{8/12}\rfloor}$}
\\
\cline{3-4}\cline{5-6}\cline{7-8}\cline{9-10}\cline{11-12}\cline{13-14}
$N$ & \multicolumn{1}{c|}{}&  \multicolumn{8}{c||}{$m_N=\lfloor \alpha N/k_N\rfloor$}  
 & \multicolumn{4}{c}{$m_N=\lfloor \alpha N^{4/3}/k_N\rfloor$}

 \\
\cline{3-4}\cline{5-6}\cline{7-8}\cline{9-10}\cline{11-12}\cline{13-14}
 & \multicolumn{1}{c|}{}& \multicolumn{1}{c}{$\hat\theta_1$}& \multicolumn{1}{c|}{$\hat\theta_2$} &  \multicolumn{1}{c}{$\hat\theta_1$} & \multicolumn{1}{c|}{$\hat\theta_2$} &  \multicolumn{1}{c}{$\hat\theta_1$} & \multicolumn{1}{c|}{$\hat\theta_2$} & \multicolumn{1}{c}{$\hat\theta_1$} & \multicolumn{1}{c||}{$\hat\theta_2$}  
 & \multicolumn{1}{c}{$\hat\theta_1$} & \multicolumn{1}{c|}{$\hat\theta_2$}
  & \multicolumn{1}{c}{$\hat\theta_1$} & \multicolumn{1}{c}{$\hat\theta_2$}
 \\
\hline
&BIAS&-0.15  &27.32&-0.28  &10.87&-0.21  &5.63 &-0.37  &2.93 &-0.36  &27.64 &-0.10   &2.84 \\
 & SD&9.75   &16.44&7.37   &10.11&7.42   &9.32&7.18   &8.73&6.33   &9.39&5.21   &6.41  \\
&  RMSE &9.75   &31.88 &7.37   &14.85&7.42   &10.89&7.19   &9.21&6.34   &29.19 &5.21   &7.01\\
&ASD&4.92   &5.88 &4.92   &5.88&4.92   &5.88&4.92   &5.88&4.92   &5.88&4.92   &5.88\\
&{$\alpha$-adjusted ASD}&6.96   &8.32 &6.96   &8.32&6.96   &8.32  &6.96   &8.32&6.96   &8.32&6.96   &8.32\\
2,000& SSE&8.06   &13.49&5.34   &7.08 &5.03   &6.25&4.80    &5.93&7.39   &12.15&4.83   &5.92\\
&$\alpha$-adjusted  SSE &11.39  &19.07&7.55   &10.02&7.12   &8.84&6.79   &8.38&10.45  &17.18&6.84   &8.37 \\
&CP (\%)&85.5&  25.7&82.1&  61.9&79.4&  72.7&80.3&77.2&95.9& 25.5&92.4& 89.8\\
&$\alpha$-adjusted CP (\%)&96.3&  55.2&94.6& 83.0&92.0&  89.3&91.8&91.0&99.8& 71.7&99.0&98.3\\
& MEMORY&\multicolumn{2}{c|}{0.55}&\multicolumn{2}{c|}{1.06}&\multicolumn{2}{c|}{2.02}&\multicolumn{2}{c||}{3.79} &\multicolumn{2}{c|}{0.55} &\multicolumn{2}{c}{3.79}    \\
\hline
&BIAS&0.08   &10.31 &-0.07  &4.34&-0.02  &1.94 &0.06   &0.99  &-0.01  &10.12&0.03   &0.87 \\
 & SD&3.30    &4.40&3.14   &3.98&3.11   &3.76&3.10    &3.70 &2.32   &3.06&2.19   &2.73 \\
&  RMSE &3.30    &11.21 &3.15   &5.88 &3.11   &4.23 &3.10    &3.83&2.32   &10.57&2.19   &2.86 \\
&ASD&2.20    &2.63&2.20    &2.63&2.20    &2.63&2.20    &2.63&2.20    &2.63&2.20    &2.63 \\
&$\alpha$-adjusted ASD &3.11   &3.72&3.11   &3.72&3.11   &3.72&3.11   &3.72&3.11   &3.72&3.11   &3.72 \\
{10,000}& SSE&2.39   &3.15 &2.27   &2.80&2.22   &2.68&2.17   &2.61&2.39   &3.14 &2.17   &2.60\\
&$\alpha$-adjusted SSE &3.39   &4.46&3.21   &3.96 &3.14   &3.79&3.07   &3.69&3.38   &4.44&3.06   &3.68\\
&CP (\%)&84.9&16.1&84.3&61.2&82.2&78.4&80.1&81.8 &96.0&9.0&95.0&92.9     \\
&$\alpha$-adjusted CP (\%)&94.7& 34.2&95.8& 82.0&94.5&92.4&93.2&92.2&99.5&31.1&99.3& 99.1    \\
& MEMORY&\multicolumn{2}{c|}{1.10}&\multicolumn{2}{c|}{2.40}&\multicolumn{2}{c|}{5.16}&\multicolumn{2}{c||}{11.14}&\multicolumn{2}{c|}{1.10} &\multicolumn{2}{c}{11.14}    \\
\hline
&BIAS&-0.03  &4.75&-0.06  &1.85&-0.04  &0.78 &-0.05  &0.21&-0.04  &4.70 &-0.04  &0.23 \\
 & SD &1.38   &1.76 &1.36   &1.70 &1.37   &1.64&1.35   &1.64&0.99   &1.27&0.97   &1.20\\
&  RMSE &1.38   &5.07 &1.36   &2.51 &1.37   &1.82&1.35   &1.65&1.00    &4.87&0.97   &1.22\\
&ASD&0.98   &1.18 &0.98   &1.18&0.98   &1.18 &0.98   &1.18&0.98   &1.18&0.98   &1.18\\
&$\alpha$-adjusted ASD &1.39   &1.67 &1.39   &1.67&1.39   &1.67&1.39   &1.67&1.39   &1.67&1.39   &1.67\\
{50,000}& SSE &1.02   &1.28&1.00    &1.21&0.99   &1.18 &0.97   &1.16&1.02   &1.27&0.97   &1.17\\
&$\alpha$-adjusted SSE  &1.45   &1.80&1.41   &1.71&1.40    &1.68 &1.37   &1.65&1.45   &1.80&1.38   &1.65\\
&CP (\%)&85.1& 8.8&83.7& 62.6&83.2& 80.8&83.5& 82.5&96.0& 3.5&94.5& 94.5\\
&$\alpha$-adjusted CP (\%)&95.8& 24.3&95.7& 81.6&96.0& 93.3&95.2& 95.0&99.8& 17.1&99.8& 99.4   \\
& MEMORY &\multicolumn{2}{c|}{2.16}  &\multicolumn{2}{c|}{5.35}&\multicolumn{2}{c|}{13.20}&\multicolumn{2}{c||}{32.57} &\multicolumn{2}{c|}{2.16} &\multicolumn{2}{c}{32.57}       \\
\hline
\end{tabular}}
\end{center}

\end{table}

\begin{table}[htbp!]
\caption{The average bias (BIAS), standard deviation (SD), root mean
squared error (RMSE), asymptotic standard deviation (ASD),  average subbagging standard error (SSE), empirical coverage probability (CP), and memory usage ({\color{black}MEMORY, in KiB}) of the subbagging estimates. Values of BIAS, SD, RMSE, ASD and SSE in this table are 100 times  their original values. Only the settings  in bold satisfy the requirement of Algorithm \ref{al:2}.}  \label{tb:3}

\begin{center}
\scalebox{0.66}{
\begin{tabular}{c|c|rr|rr|rr|rr||rr|rr}
   \hline%full sample {\color{black}estimator}

\multicolumn{14}{c}{Algorithm \ref{al:2} with $\hat \theta_{k_N,m_N}^{(bc2)}=m_N^{-1}\sum_{s\in\mathcal{S}}\hat\theta_{k_N,s}^{(bc2)}$ {\color{black} for Logistic Regression}}   \\
  \hline
   && \multicolumn{12}{c}{$\alpha=1$}\\  
\cline{3-4}\cline{5-6}\cline{7-8}\cline{9-10}\cline{11-12}\cline{13-14} 
 && \multicolumn{2}{c|}{$\boldsymbol {k_N=\lfloor N^{5/12}\rfloor}$}    & \multicolumn{2}{c|}{$\boldsymbol {k_N=\lfloor N^{6/12}\rfloor}$}      &
\multicolumn{2}{c|}{${k_N=\lfloor N^{7/12}\rfloor}$}   & \multicolumn{2}{c||}{${ k_N=\lfloor N^{8/12}\rfloor}$}&  \multicolumn{2}{c|}{$\boldsymbol{k_N=\lfloor N^{5/12}\rfloor}$}
&  \multicolumn{2}{c}{${k_N=\lfloor N^{8/12}\rfloor}$}
\\
\cline{3-4}\cline{5-6}\cline{7-8}\cline{9-10}\cline{11-12}\cline{13-14}
$N$& \multicolumn{1}{c|}{}&  \multicolumn{8}{c||}{$m_N=\lfloor \alpha N/k_N\rfloor$}  
 & \multicolumn{4}{c}{$m_N=\lfloor \alpha N^{4/3}/k_N\rfloor$}

 \\
\cline{3-4}\cline{5-6}\cline{7-8}\cline{9-10}\cline{11-12}\cline{13-14}
\cline{3-4}\cline{5-6}\cline{7-8}\cline{9-10}\cline{11-12}\cline{13-14}
 & \multicolumn{1}{c|}{}& \multicolumn{1}{c}{$\hat\theta_1$}& \multicolumn{1}{c|}{$\hat\theta_2$} &  \multicolumn{1}{c}{$\hat\theta_1$} & \multicolumn{1}{c|}{$\hat\theta_2$} &  \multicolumn{1}{c}{$\hat\theta_1$} & \multicolumn{1}{c|}{$\hat\theta_2$} & \multicolumn{1}{c}{$\hat\theta_1$} & \multicolumn{1}{c||}{$\hat\theta_2$}  
 & \multicolumn{1}{c}{$\hat\theta_1$} & \multicolumn{1}{c|}{$\hat\theta_2$}
  & \multicolumn{1}{c}{$\hat\theta_1$} & \multicolumn{1}{c}{$\hat\theta_2$}
 \\
\hline
&BIAS &11.54  &29.33 &-0.36  &6.82&-0.21  &1.45&-0.37  &0.53&3.86   &11.97&-0.10   &0.44\\
 & SD&6.92   &20.44&8.12   &16.75&7.33   &9.24&7.12   &8.63&4.07   &14.23&5.16   &6.33\\
&  RMSE&13.45  &35.75&8.13   &18.08&7.33   &9.35&7.13   &8.64 &5.61   &18.60&5.17   &6.34\\
&ASD&4.93   &5.92&4.92   &5.88&4.92   &5.88&4.92   &5.88&4.93   &5.90&4.92   &5.88\\
&$\alpha$-adjusted ASD&6.97   &8.37&6.96   &8.32&6.96   &8.32&6.96   &8.32&6.97   &8.35&6.96   &8.32\\
2,000& SSE&6.98   &16.39&6.36   &15.00&4.98   &6.23&4.76   &5.86&4.81   &12.08&4.80    &5.85\\
&$\alpha$-adjusted SSE&9.88   &23.17&8.99   &21.21&7.05   &8.80&6.74   &8.29&6.80    &17.09&6.78   &8.28\\
&CP (\%)&61.5&52.9&83.0&80.2&79.9&80.8&80.3&79.4&96.7&83.3&92.5&92.2\\
&$\alpha$-adjusted CP (\%)&89.4&87.8&94.6&93.3&92.0&92.5&91.7&91.9&100.0&98.7&99.0&98.6\\
& MEMORY&\multicolumn{2}{c|}{0.55}&\multicolumn{2}{c|}{1.06} &\multicolumn{2}{c|}{2.02}&\multicolumn{2}{c||}{3.79} &\multicolumn{2}{c|}{0.55} &\multicolumn{2}{c}{3.79}     \\
\hline
&BIAS&0.08   &5.66&-0.07  &0.71&-0.02  &0.14&0.06   &0.14&-0.09&5.49 &0.03   &0.02\\
 & SD&3.75   &7.67&3.12   &3.94&3.09   &3.73&3.09   &3.68 &2.97 &3.93&2.19   &2.71\\
&  RMSE&3.75   &9.53&3.12   &4.00&3.09   &3.73 &3.09   &3.69&2.98 &6.75&2.19   &2.71\\
&ASD&2.20    &2.63&2.20    &2.63&2.20    &2.63&2.20    &2.63&2.20    &2.63&2.20    &2.63\\
&$\alpha$-adjusted ASD&3.11   &3.72&3.11   &3.72&3.11   &3.72&3.11   &3.72&3.11   &3.72&3.11   &3.72\\
10,000& SSE&2.99   &6.95&2.25   &2.77&2.20    &2.65&2.16   &2.60 &3.09 &4.91&2.16   &2.59\\
&$\alpha$-adjusted SSE&4.22   &9.82&3.18   &3.92&3.12   &3.75&3.06   &3.67&4.38 &6.95&3.05   &3.66\\
&CP (\%)&85.0&65.2&83.5&82.0&82.2&82.6&79.9&80.5&96.7&81.1&94.9&93.8\\
&$\alpha$-adjusted CP (\%)&94.9&87.3&95.7&94.0&94.6&94.7&93.2&93.4&99.6&96.9&99.3&99.2\\
& MEMORY  &\multicolumn{2}{c|}{1.10}&\multicolumn{2}{c|}{2.40}&\multicolumn{2}{c|}{5.16}&\multicolumn{2}{c||}{11.14} &\multicolumn{2}{c|}{1.10} &\multicolumn{2}{c}{32.57}   \\
\hline
&BIAS&-0.03  &0.81&-0.06  &0.11 &-0.04  &0.06&-0.05  &-0.08&-0.04&0.76&-0.04&-0.06\\
 & SD&1.37   &1.74&1.35   &1.68&1.37   &1.63&1.34   &1.64&0.98 &1.25&0.97 &1.20\\
&  RMSE&1.37   &1.92&1.35   &1.69&1.37   &1.63&1.35   &1.64&0.99 &1.47&0.97 &1.20\\
&ASD &0.98   &1.18&0.98   &1.18&0.98   &1.18&0.98   &1.18&0.98 &1.18&0.98 &1.18\\
&$\alpha$-adjusted ASD&1.39   &1.67 &1.39   &1.67&1.39   &1.67 &1.39   &1.67&1.39 &1.67&1.39 &1.67\\
50,000& SSE&1.01   &1.27&0.99   &1.20&0.99   &1.18&0.97   &1.16&1.01 &1.27&0.97 &1.17\\
&$\alpha$-adjusted SSE&1.43   &1.79&1.40    &1.70&1.39   &1.67&1.37   &1.64&1.43 &1.79&1.37 &1.65\\
&CP (\%)&85.0&79.4&83.7&83.9&83.1&83.3&83.5&82.7&95.9&91.0&94.5&94.2\\
&$\alpha$-adjusted CP (\%)&95.7&93.1&95.7&94.6&96.0&94.9&95.2&94.5&99.8&98.5&99.8&99.5\\
& MEMORY &\multicolumn{2}{c|}{2.16} &\multicolumn{2}{c|}{5.35} &\multicolumn{2}{c|}{13.20} &\multicolumn{2}{c||}{32.57} &\multicolumn{2}{c|}{2.16}&\multicolumn{2}{c}{32.57}   \\
\hline
\end{tabular}}
\end{center}

\end{table}

\begin{table}[htbp!]
\caption{The average bias (BIAS), standard deviation (SD), root mean
squared error (RMSE), asymptotic standard deviation (ASD),  average subbagging standard error (SSE), empirical coverage probability (CP), and memory usage ({\color{black}MEMORY, in KiB}) of the subbagging estimates. Values of BIAS, SD, RMSE, ASD and SSE in this table are 100 times  their original values. Only the settings  in bold satisfy the requirement of Algorithm  \ref{al:2}.}  \label{tb:4}
\begin{center}
\scalebox{0.66}{
\begin{tabular}{c|c|rr|rr|rr|rr||rr|rr}
   \hline%full sample {\color{black}estimator}

\multicolumn{14}{c}{Algorithm \ref{al:2} with $\hat \theta_{k_N,m_N}^{(bc3)}=m_N^{-1}\sum_{s\in\mathcal{S}}\hat\theta_{k_N,s}^{(bc3)}$ {\color{black} for Logistic Regression}}   \\
  \hline
   && \multicolumn{12}{c}{$\alpha=1$}\\  
\cline{3-4}\cline{5-6}\cline{7-8}\cline{9-10}\cline{11-12}\cline{13-14} 
 && \multicolumn{2}{c|}{$\boldsymbol {k_N=\lfloor N^{5/12}\rfloor}$}    & \multicolumn{2}{c|}{$\boldsymbol {k_N=\lfloor N^{6/12}\rfloor}$}      &
\multicolumn{2}{c|}{${k_N=\lfloor N^{7/12}\rfloor}$}   & \multicolumn{2}{c||}{${ k_N=\lfloor N^{8/12}\rfloor}$}&  \multicolumn{2}{c|}{$\boldsymbol{k_N=\lfloor N^{5/12}\rfloor}$}
&  \multicolumn{2}{c}{${k_N=\lfloor N^{8/12}\rfloor}$}
\\
\cline{3-4}\cline{5-6}\cline{7-8}\cline{9-10}\cline{11-12}\cline{13-14}
$N$& \multicolumn{1}{c|}{}&  \multicolumn{8}{c||}{$m_N=\lfloor \alpha N/k_N\rfloor$}  
 & \multicolumn{4}{c}{$m_N=\lfloor \alpha N^{4/3}/k_N\rfloor$}

 \\
\cline{3-4}\cline{5-6}\cline{7-8}\cline{9-10}\cline{11-12}\cline{13-14}
\cline{3-4}\cline{5-6}\cline{7-8}\cline{9-10}\cline{11-12}\cline{13-14}
 & \multicolumn{1}{c|}{}& \multicolumn{1}{c}{$\hat\theta_1$}& \multicolumn{1}{c|}{$\hat\theta_2$} &  \multicolumn{1}{c}{$\hat\theta_1$} & \multicolumn{1}{c|}{$\hat\theta_2$} &  \multicolumn{1}{c}{$\hat\theta_1$} & \multicolumn{1}{c|}{$\hat\theta_2$} & \multicolumn{1}{c}{$\hat\theta_1$} & \multicolumn{1}{c||}{$\hat\theta_2$}  
 & \multicolumn{1}{c}{$\hat\theta_1$} & \multicolumn{1}{c|}{$\hat\theta_2$}
  & \multicolumn{1}{c}{$\hat\theta_1$} & \multicolumn{1}{c}{$\hat\theta_2$}
 \\
\hline
&BIAS&10.62  &10.96 &-0.30   &5.74&-0.21  &1.69&-0.37  &0.61&4.53   &12.71&-0.10   &0.51\\
 & SD&5.70    &9.60 &7.41   &10.54 &7.33   &9.25&7.12   &8.63&3.74   &7.32 &5.17   &6.33\\
&  RMSE&12.05  &14.56&7.41   &12.0&7.34   &9.40&7.13   &8.65&5.87   &14.67 &5.17   &6.35\\
&ASD&4.93   &5.92&4.92   &5.88&4.92   &5.88&4.92   &5.88&4.92   &5.88&4.92   &5.88\\
&$\alpha$-adjusted ASD&6.97   &8.37&6.96   &8.32 &6.96   &8.32&6.96   &8.32&6.96   &8.32&6.96   &8.32\\
2,000& SSE&5.91   &7.55 &5.41   &7.46&4.99   &6.24&4.76   &5.86 &6.06   &7.65&4.80    &5.85\\
&$\alpha$-adjusted SSE&8.35   &10.67 &7.65   &10.56&7.05   &8.82&6.74   &8.29&8.57   &10.82&6.78   &8.28\\
&CP (\%)&56.9&66.3&82.4&78.0&79.8&80.5&80.3&79.1&94.6&63.3&92.5&92.2\\
&$\alpha$-adjusted CP (\%)&84.6&86.7&94.4&92.7&92.0&92.3&91.7&91.9&99.3&86.7&99.0&98.6\\
& MEMORY&\multicolumn{2}{c|}{0.55}&\multicolumn{2}{c|}{1.06}&\multicolumn{2}{c|}{2.02} &\multicolumn{2}{c||}{3.79} &\multicolumn{2}{c|}{0.55}&\multicolumn{2}{c}{3.79}  \\
\hline
&BIAS&0.09   &5.18&-0.07  &0.88&-0.02  &0.18 &0.06   &0.15&-0.01  &4.99&0.03   &0.03\\
 & SD&3.31   &4.54&3.12   &3.94&3.09   &3.73&3.09   &3.68&2.32   &3.14&2.19   &2.71\\
&  RMSE&3.32   &6.88&3.12   &4.04&3.09   &3.73&3.09   &3.69&2.32   &5.90&2.19   &2.71\\
&ASD&2.20    &2.63 &2.20    &2.63&2.20    &2.63&2.20    &2.63&2.20    &2.63&2.20    &2.63\\
&$\alpha$-adjusted ASD&3.11   &3.72&3.11   &3.72&3.11   &3.72&3.11   &3.72 &3.11   &3.72&3.11   &3.72\\
10,000& SSE&2.42   &3.30&2.25   &2.78&2.2    &2.65&2.16   &2.60 &2.42   &3.30&2.16   &2.59\\
&$\alpha$-adjusted SSE &3.42   &4.67&3.18   &3.92&3.12   &3.75&3.06   &3.67 &3.42   &4.66&3.05   &3.66\\
&CP (\%)&85.1&60.6&83.4&82.1&82.3&82.5&79.9&80.7&95.9&69.2&94.9&93.8\\
&$\alpha$-adjusted CP (\%)&94.9&82.2&95.7&94.3&94.6&94.9&93.2&93.3&99.5&91.7&99.3&99.2\\
& MEMORY&\multicolumn{2}{c|}{1.10}&\multicolumn{2}{c|}{2.40}&\multicolumn{2}{c|}{5.16}&\multicolumn{2}{c||}{11.14}&\multicolumn{2}{c|}{1.10} &\multicolumn{2}{c}{11.14}   \\
\hline
&BIAS &-0.03  &1.01&-0.06  &0.15&-0.04  &0.07&-0.05  &-0.08&-0.04&0.97 &-0.04&-0.06\\
 & SD&1.37   &1.74 &1.35   &1.68&1.37   &1.63&1.34   &1.64&0.98 &1.26&0.97 &1.20\\
&  RMSE&1.37   &2.01&1.35   &1.69&1.37   &1.64&1.35   &1.64&0.99 &1.58&0.97 &1.20\\
&ASD&0.98   &1.18&0.98   &1.18&0.98   &1.18&0.98   &1.18&0.98 &1.18&0.98 &1.18\\
&$\alpha$-adjusted ASD&1.39   &1.67&1.39   &1.67&1.39   &1.67&1.39   &1.67 &1.39 &1.67&1.39 &1.67\\
50,000& SSE&1.01   &1.27&0.99   &1.20&0.99   &1.18&0.97   &1.16&1.01 &1.27&0.97 &1.17\\
&$\alpha$-adjusted SSE&1.43   &1.79&1.40    &1.70&1.39   &1.67&1.37   &1.64&1.43 &1.79&1.36 &1.66\\
&CP (\%)&85.0&77.7&83.7&83.7&83.1&83.3&83.5&82.7&95.9&88.6&94.5&94.2\\
&$\alpha$-adjusted CP (\%)&95.7&91.7&95.7&94.6&96.0&95.0&95.2&94.5&99.8&98.0&99.9&99.6\\
& MEMORY&\multicolumn{2}{c|}{2.16} &\multicolumn{2}{c|}{5.35}&\multicolumn{2}{c|}{13.20}&\multicolumn{2}{c||}{32.57} &\multicolumn{2}{c|}{2.16} &\multicolumn{2}{c}{32.57}    \\
\hline
\end{tabular}}
\end{center}
\end{table}

\begin{table}[htbp!]
\caption{The results of fitting the logistic  regression model with variables Intercept, Year,  CRSDepTime and ActualElapsedTime. The SE is either the standard error for the full sample estimate or the $\alpha$-adjusted SSE for the subbagging estimate. 
\label{tb:5}}
\begin{center}
\scalebox{0.78}{\begin{tabular}{c|r||c|r|r|r}
\hline & \multicolumn{1}{c||}{\makecell[cb]{Full Sample \\ Estimate  }}&&\multicolumn{1}{c|}{ \makecell[cb]{Algorithm \ref{al:1}\\$\hat \theta_{k_N,m_N}$ }} &  \multicolumn{1}{c|}{\makecell[cb]{Algorithm \ref{al:2}\\$\hat \theta_{k_N,m_N}^{(bc2)}$ }}&\multicolumn{1}{c}{\makecell[cb]{Algorithm \ref{al:2}  \\$\hat \theta_{k_N,m_N}^{(bc3)}$ }}\\

\hline & \multicolumn{1}{c||}{\multirow{2}{*}{\footnotesize$N$=118,914,459}} && \multicolumn{1}{c|}{\footnotesize$ k_N=\lfloor N^{1/2+1/1000}\rfloor{\color{black}= 11,109}$} & \multicolumn{2}{c}{\footnotesize$ k_N=\lfloor N^{1/3+1/1000}\rfloor{\color{black}=500}$}   \\
\cline{4-6}&&& \multicolumn{3}{c}{\footnotesize$m_N=\lfloor \alpha N/k_N\rfloor$}\\

\hline Intercept    &  $ 52.7403$&&$ 52.7377$ &$ 52.4029$ &$52.2312$ \\
SE &(0.06)&&(0.65)&(0.74)&(0.23)\\
\cline{1-2}\cline{4-6} Year& $  -26.8800$  && $ -26.7191$ &$ -26.7262$ &$-26.4874$ \\
SE &(0.03)&&(0.32)&(0.38)&(0.12)\\
\cline{1-2}\cline{4-6} CRSDepTime& $ 0.3780$  && $ 0.3930$ &$0.3906$ &$ 0.4020$ \\
SE ($\times 10^{-3}$)&(0.39)&$\alpha=0.01$&(3.85)&(4.06)&(1.08)\\
\cline{1-2}\cline{4-6} ActualElapsedTime& $3.2600 $  && $3.2540$ &$3.2249$ &$3.3156$ \\
SE ($\times 10^{-3}$)&(2.81)&&(26.36)&(27.77)&(17.85)\\
\cline{1-2}\cline{4-6} Loading Time (in minutes) &{\color{black}27.7} &&{\color{black}0.5}&{\color{black}6.5} &{\color{black}6.3}\\
 Estimation Time (in minutes) &{\color{black}20.9} &&{\color{black}0.4}&{\color{black}2.0} &{\color{black}3.3}\\
 SE Time  (in seconds) &1.7 &&0.0003&0.0007&0.0007\\
\cline{1-2}\cline{4-6} {\color{black}MEMORY (in KiB)}  &12686071.1  &&2666.1 &120.0 &120.0 \\
\hline
\hline Intercept    & & &$ 52.6310 $ &$ 52.6287$ &$  52.3798 $ \\
SE &&&(0.11)&(0.18)&(0.08)\\
\cline{1-1}\cline{4-6}
Year            &   && $ -26.8360$ &$-26.8343 $ &$-26.5621  $ \\
SE &&&(0.05)&(0.09)&(0.04)\\
\cline{1-1}\cline{4-6}
CRSDepTime             &   && $ 0.3908$ &$ 0.3912$ &$0.4021  $ \\
SE ($\times 10^{-3}$)&&$\alpha=0.2$&(0.98)&(0.98)&(0.29)\\
\cline{1-1}\cline{4-6} ActualElapsedTime  & & &$ 3.2742$ &$ 3.2614 $ &$ 3.3259 $ \\
SE ($\times 10^{-3}$)&&&(6.77)&(7.06)&(5.80)\\
\cline{1-1}\cline{4-6} Loading Time (in minutes) & & &{\color{black}9.3}&{\color{black}134.1} &{\color{black}134.2} \\
 Estimation Time (in minutes) &&  &{\color{black}8.2} &{\color{black}41.1} &{\color{black}73.4}\\
 SE Time  (in seconds) & && {\color{black}0.0013}&{\color{black}0.0170}&{\color{black}0.0150}\\
\cline{1-1}\cline{4-6} {\color{black}MEMORY (in KiB)} && &2666.1 &120.0 &120.0 \\
\hline 
\end{tabular}}
\end{center}
\end{table}

\end{document}